\documentclass[sigconf,nonacm]{acmart}

\AtBeginDocument{%
  }

\usepackage{amsmath}
\usepackage{amsthm}
\usepackage{multirow}
\usepackage{xspace}
\usepackage{xurl}
\usepackage[most]{tcolorbox}
\usepackage[ruled,vlined,linesnumbered]{algorithm2e}

\newtcolorbox{promptbox}[1]{%
  enhanced, breakable, sharp corners, colback=black!3, colframe=black!45,
  boxrule=0.5pt, left=4pt, right=4pt, top=4pt, bottom=4pt,
  fonttitle=\bfseries\footnotesize, coltitle=white, colbacktitle=black!60,
  title={#1}, before skip=6pt, after skip=6pt}

\newcommand{\prompttext}[1]{%
  {\ttfamily\fontsize{6.6}{7.6}\selectfont\raggedright #1\par}}
\newcommand{\promptlabel}[1]{{\ttfamily\fontsize{6.6}{7.6}\selectfont\bfseries #1\par}}

\newcommand{\method}{LEDGER\xspace}
\newcommand{\methodlong}{LLM-Evidence-Driven Graph-Encoded Clustering-Friendly Representations for Short Texts\xspace}
\newcommand{\KL}{\mathrm{KL}}
\newcommand{\Beta}{\mathrm{Beta}}
\newcommand{\softp}{\mathrm{softplus}}

\newcommand{\zk}{\textcolor{black}}  

\title[\method: Graph-Encoded Representations for Short Texts]{\method: \methodlong}

\author{Jie Wang}
\affiliation{%
  \department{School of Mathematics and Statistics}
  \department{MOE Key Laboratory for Complexity Science in Aerospace}
  \institution{Northwestern Polytechnical University}
  \city{Xi'an}
  \country{China}}
\email{dogtwice@mail.nwpu.edu.cn}

\author{Kuang Zhou}
\authornote{Corresponding author. (\href{mailto:kzhoumath@nwpu.edu.cn}{kzhoumath@nwpu.edu.cn}).}
\affiliation{%
  \department{School of Mathematics and Statistics}
  \department{MOE Key Laboratory for Complexity Science in Aerospace}
  \institution{Northwestern Polytechnical University}
  \city{Xi'an}
  \country{China}}
\email{kzhoumath@nwpu.edu.cn}

\renewcommand{\shortauthors}{Wang and Zhou}

\begin{abstract}
\zk{Short-text clustering is important for analyzing user-generated content in social networks and online services, supporting applications such as topic discovery and user intent analysis. However, the desired grouping depends on an application-specific clustering criterion that general-purpose text embeddings may not capture.} 
Large language models (LLMs) can apply such a criterion to a pair of texts without human annotation, and existing methods use their responses as generated descriptions, binary decisions or selected neighbours. These discrete outputs discard how strongly the LLM supports a pair and how many judgments stand behind it. Each judgment is also used alone or within the neighbourhood of a single text, so the relations that link judgments across the corpus remain unused. In this paper, we propose \method, an LLM-guided graph representation learning framework for short-text clustering. \method aggregates the repeated numerical scores of each text pair into an accumulated score and an observation count, which define an evidence target for every queried pair. The pairs whose scores reach a threshold form a relation graph, over which a graph encoder propagates text features with a propagation range learned jointly with the encoder. A pairwise predictor trained to match the evidence targets shapes the learned representation, and a distilled student encoder extends it to unqueried texts without further LLM requests. Extensive experiments on seven datasets with four frozen text embeddings show that \method achieves the highest mean clustering accuracy and NMI against five comparison methods, and ablations show that the graph, the learned propagation range and the evidence target each improve performance.

\end{abstract}


\begin{document}

\maketitle

\section{Introduction}
\label{sec:intro}

\zk{Online communities and interactive services generate large collections of
short texts, including discussion titles, customer requests, and assistant
commands. Organizing this content supports information access and the
discovery of recurring topics and user intents \citep{chen-etal-2022-intent}.} 
A clusterer groups texts by their proximity in a representation space, so the partition it produces is determined by the representation. In practice, this representation is commonly produced by a general-purpose text embedding model, which encodes each text as a vector according to the semantic similarity learned from large corpora. The categories of a dataset, however, are defined by its own clustering criterion, which this general similarity need not follow. For instance, such a model places two requests about the same bank card close together, which fits a criterion that groups texts by topic but conflicts with one that separates them by intent. Consequently, no single general-purpose representation suits the criteria of all datasets. The criterion of each dataset is not contained in the texts themselves, and it has to be provided as supervision.

Semi-supervised clustering provides this supervision through labelled examples, with which the text encoder is fine-tuned so that the representation reflects the criterion \citep{zhang2021deepaligned,zhang-etal-2022-new}. Obtaining these labels, however, requires domain knowledge and substantial human annotation \citep{kumar-etal-2022-intent}, which makes such supervision costly at scale. What the annotator contributes is the application of the criterion, namely deciding which texts belong to the same category. An LLM given the criterion in natural language can make this decision for a pair of texts, and therefore offers a more accessible source of supervision without human annotation. Following the same route, LLM supervision can be brought into the representation on which the clusterer operates, which raises two design questions in representation learning, the form that the supervision takes and how it is organized into that representation.

Existing LLM-guided short-text clustering methods can be examined through these two questions. According to what the LLM is asked to produce, they fall into three families. Generated-text methods ask the LLM to write a short label or description for each text and cluster the encoded descriptions \citep{de-raedt-etal-2023-idas,viswanathan-etal-2024-large}. Pairwise and triplet methods instead query relations between texts. They show the LLM two texts and ask whether they share a category, or an anchor and two candidates and ask which candidate is closer to the anchor. The answers are then either used to fine-tune the embedder \citep{zhang-etal-2023-clusterllm,fan-etal-2024-lanid} or imposed as must-link and cannot-link constraints in the clustering objective \citep{wang-etal-2025-cequel}. Finally, neighbour-selection methods retrieve candidates near a target text, ask the LLM to select those that share the category of the target, and pool the target with the selected texts into its representation \citep{lin-etal-2025-spill,lin-etal-2026-llms}. Despite their different queries, the three families answer the first question in the same way. In each family, an LLM response enters the pipeline as a discrete output, namely a description, a binary decision, or a set of selected candidates. Such outputs do not record how strongly the LLM supports two texts sharing a category. Yet this strength is directly available. An LLM can state it as a numerical score for a pair of texts, and verbalized confidence scores elicited in this way have been shown to carry information about whether the judgment is correct \citep{tian-etal-2023-just}. Beyond the score, the number of observations indicates how much evidence stands behind it when a pair is judged more than once, as when each of its two texts serves as a target. Existing methods, however, retain neither quantity.

The second question concerns how the LLM judgments enter the representation on which the clusterer operates. Generated-text methods act on each text individually, generating and encoding the description of each text separately. Among pairwise and triplet methods, embedder fine-tuning treats each triplet response as a separate loss term, while constraint-based clustering adds each judgment to the clustering objective as an independent constraint. Neighbour-selection methods go beyond single texts, but they pool each target only with its own selected texts under a prespecified rule. Across the three families, each judgment is thus used in isolation or within the neighbourhood of a single target. The judgments, however, are not isolated from one another. A text takes part in many judgments, so judgments that share a text are linked. For example, if one judgment places texts A and B in the same category and another places B and C together, the two judgments jointly relate A to C, although neither involves both texts. Linking all judgments in this way yields a relation graph over the corpus, with texts as nodes and judged relations as edges. In this view, existing methods use the graph edge by edge or within the one-hop neighbourhood of a target, and the effect of that neighbourhood on the representation is fixed in advance. Relations beyond one hop are therefore either left unused or exert an influence that cannot be learned from the supervision.

We introduce \method, an LLM-guided graph representation learning framework for short-text clustering. For the form of supervision, \method models LLM judgments as pairwise evidence. Specifically, the LLM assigns a numerical score to each retrieved pair for whether the two texts share a category, and the repeated observations of a pair are aggregated into an accumulated score and an observation count. The two quantities define a Beta distribution for the pair, whose mean follows the accumulated score and whose concentration grows with the observation count, and this Beta distribution serves as the evidence target that supervises every queried pair. The same judgments also construct the relation graph, in which each relation whose score reaches an acceptance threshold becomes an edge. For the organization of supervision, a graph encoder propagates text features over this relation graph. Following adaptive graph filters \citep{chien2021gprgnn}, the range of propagation is learned jointly with the encoder rather than fixed in advance, and a pairwise predictor on the learned representations is trained to match the evidence target of each queried pair. The learned representation is then passed to a downstream clusterer. Beyond the queried corpus, representation distillation extends the learned space to unqueried texts from the same distribution, and the block budget offers several request operating points for deployment.

We make four contributions:
\begin{itemize}
  \item \textbf{LLM-guided graph representation framework.} We formulate
    short-text clustering as LLM-guided graph representation learning. The
    framework organizes the relations obtained from the LLM into a graph over
    the corpus and learns how far relational context propagates in the
    clustering representation.
  \item \textbf{Evidence modeling of LLM judgments.} We model each LLM
    judgment as pairwise evidence. The numerical score records the strength of
    support and the observation count records the amount of evidence; together they define the evidence target that supervises every queried pair. The same evidence also
    determines the accepted relations in the graph.
  \item \textbf{Budget-flexible querying.} The number of LLM requests is $bN$
    for a corpus of $N$ texts, so it grows linearly with the corpus and is set
    by the block budget $b$. The judgments of a single pass serve all of
    training, and representation distillation encodes unqueried texts from
    the same distribution without further LLM requests.
  \item \textbf{Performance across budgets and corpora.} Across seven datasets
    and four frozen text embeddings, the learned representations achieve the
    strongest mean clustering performance. The block budget traces a
    request--quality curve that reaches the empirical Pareto frontier, and the
    distilled student achieves the highest mean on full-corpus extension.
\end{itemize}

\section{Related Work}
\label{sec:related}

This section reviews three bodies of work that bear on \method. We first cover short-text clustering without LLMs, which shows why the clustering criterion has to be supplied as supervision. We then cover LLM-guided methods, separating those that guide the clustering procedure from those that change the representation, since \method belongs to the latter.

\subsection{Short-Text Clustering}
Short texts are sparse, so the representation determines most of what a clusterer can recover. Early methods describe each text by its words and infer clusters from word co-occurrence with topic models \citep{yin-wang-2014-dirichlet,yan-etal-2013-biterm}, which ignore word order and semantics. Later methods represent texts with pretrained embeddings, add them to the word-based model as a second view \citep{cheng-etal-2026-nonparametric}, or reshape the embedding with contrastive learning \citep{zhang-etal-2021-supporting}. These methods need no labels, but the clustering criterion is fixed implicitly by the representation that is built. Nothing in the data selects a criterion \citep{vonluxburg-etal-2012-clustering}: MASSIVE labels the same utterances by intent and by scenario \citep{fitzgerald-etal-2023-massive}, and no single embedding performs best across tasks \citep{muennighoff-etal-2023-mteb}. A method without labels can therefore score lower when the criterion differs from the one its representation favours, even on the same dataset.

Semi-supervised methods state the criterion explicitly through labelled examples. Pairwise constraints are enforced during clustering \citep{wagstaff2001cop,basu2004active,zhang-etal-2021-framework,ren-etal-2019-semi} and queried selectively to limit their number \citep{xiong-etal-2014-active,zhang-etal-2025-semi-supervised,wang-davidson-2010-active}, and for short texts the labelled examples fine-tune the encoder \citep{zhang2021deepaligned,zhang-etal-2022-new,lin-etal-2020-discovering,zhang-etal-2024-clustering}. Each of these routes needs an annotator to label examples or answer queries, which is costly at scale. This motivates obtaining the supervision from an LLM instead of an annotator.

\subsection{LLM-Guided Clustering Algorithms}
These methods apply LLM output to the clustering procedure and leave the representation unchanged. Some turn LLM answers into constraints. Cequel queries an LLM on selected pairs and triplets, converts the answers into must-link and cannot-link constraints, and clusters under them with a weighted constrained algorithm \citep{wang-etal-2025-cequel}, and a later method has the LLM generate sets of constraints instead of single pairs to cut the number of queries \citep{jia-etal-2026-optimized}. Others act on an initial clustering. LLMEdgeRefine forms super-points to absorb outliers and asks an LLM to reassign boundary points \citep{feng-etal-2024-llmedgerefine}, and Summaries as Centroids replaces each cluster centre with a textual summary of its members \citep{sac2026}. In all of these the LLM acts at the clustering step, so the partition still depends on the embedding the texts started from \citep{hong-etal-2025-dial,islam-2026-reasoning}.

\subsection{LLM-Guided Representation Learning for Clustering}
These methods use LLM output to change the representation that is clustered. Some methods change the text before encoding. IDAS prompts an LLM for a descriptive label of each utterance and encodes it together with the utterance \citep{de-raedt-etal-2023-idas}, and NILC adds centroids generated by an LLM and rewrites ambiguous utterances \citep{wang-etal-2026-nilc}. Others change the encoder. ClusterLLM fine-tunes a small embedder on triplets answered by an LLM and uses pairwise questions to choose the clustering granularity \citep{zhang-etal-2023-clusterllm}, and the preference-based method that fine-tunes the LLM and the embedder against each other \citep{li-etal-2025-co}. A third kind changes each vector directly. SPILL asks an LLM to select the neighbours of a target that share its intent and pools them with the target \citep{lin-etal-2025-spill}, and LeBoT builds a bag-of-texts representation that is refined iteratively by mean-pooling the selected neighbours \citep{lin-etal-2026-llms}. In these methods the LLM output enters as a generated text, a preference, a triplet choice or a selected neighbour, and the strength of each judgment is not carried into the representation \citep{yang-etal-2024-tec,an-etal-2024-loop,fan-etal-2025-ligd}. \method instead keeps the numerical score and observation count of every judged pair and learns the representation on the relation graph that the judgments form, with the contribution of each hop learned from the same supervision.

\section{Methodology}
\label{sec:method}

This section presents \method, an LLM-guided graph representation learning
framework for short-text clustering. Section~\ref{sec:method:overview}
formulates the representation-learning task and outlines the framework.
Section~\ref{sec:method:acq} models LLM judgments as pairwise evidence and
constructs a relation graph from the same judgments.
Section~\ref{sec:method:repr} then presents representation learning over this
graph under the evidence targets, and
Section~\ref{sec:method:cluster} presents the distillation that extends the
learned space to unqueried texts.

\subsection{Framework Overview}
\label{sec:method:overview}

\begin{figure*}[t]
\centering
\includegraphics[width=\textwidth]{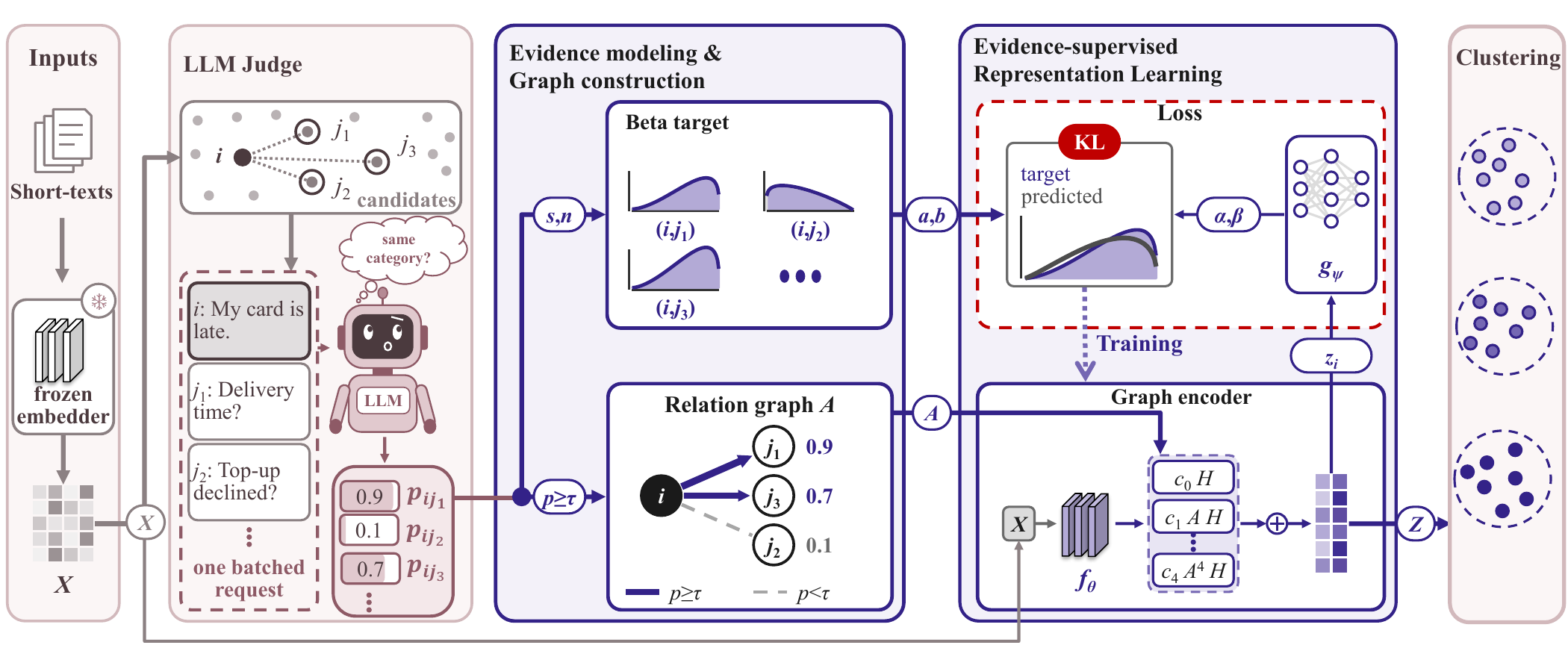}
\Description{A left-to-right architecture diagram with a worked example. The
queried texts are embedded, and each query text is judged against its retrieved
candidates by an LLM, which returns one score per candidate. The
scores split into two uses: accumulated over both query directions they form a
evidence target, and thresholded they form a relation graph. A graph
encoder with learned per-hop weights produces the representation passed to
K-Means, and a pairwise predictor is trained by KL divergence against the Beta
targets.}
\caption{Overview of \method. Each query text is judged against its $b\cdot w$
retrieved candidates in one batched request, and the judge returns a
score for each candidate. The same judgments serve two roles.
Accumulated over both query directions they form an evidence target that is retained for every judged pair, including pairs below the acceptance
threshold $\tau$; thresholded at $\tau$ they form the relation graph. A graph
encoder with learned per-hop weights yields the clustering representation $Z$,
and a pairwise predictor is trained by KL divergence to match the evidence
targets.}
\label{fig:overview}
\end{figure*}

We are given $N$ unlabelled short texts $V=\{1,\dots,N\}$, which form the
queried corpus, and their frozen text embeddings
\begin{equation}
  X=[x_1,\dots,x_N]^\top\in\mathbb{R}^{N\times d_x},
  \label{eq:input}
\end{equation}
where $d_x$ is the embedding dimension. Given a clustering criterion specified
in natural language, an LLM judge scores whether pairs of these texts share a
category. Our task is to learn a clustering representation $z_i$ for each
text that reflects the criterion through the judged pairs and can be used by a
downstream clustering procedure. Further texts from the same distribution that
receive no LLM request are to be represented in the same learned space.

The framework is organized around a single pass of LLM judgments, which
serves two roles. First, the scores and observation counts of the judged pairs
define pairwise evidence targets. Second, the relations whose scores reach an
acceptance threshold form the relation graph. Figure~\ref{fig:overview}
summarizes the framework; we detail each stage next. A graph encoder
propagates text features over this graph, and the encoder, whose parameters
include the range of propagation, is trained jointly with a pairwise predictor
on its outputs to fit the evidence targets. Texts that
receive no LLM request, however, have no judgments and hence no place in the
relation graph. To represent them, a student encoder distilled from the graph
encoder maps their frozen embeddings into the learned representation space.

\subsection{LLM Evidence Modeling and Relation Graph Construction}
\label{sec:method:acq}

\method asks the LLM for a numerical score for each judged pair, which records
how strongly the LLM supports the pair belonging to one category. Since the number of text
pairs grows quadratically with corpus size, each text is judged against a
bounded set of retrieved candidates under an adjustable block budget.
Specifically, for each query text $i$, we rank the other texts by cosine
similarity in the frozen embedding space and divide the top $bw$ candidates by
rank into $b$ disjoint blocks of size $w$. Each block forms one batched LLM
request, so a pass over the corpus issues $bN$ requests and the block budget
$b$ sets the request cost. For every candidate $j$ in a block, the judge
returns a score $p_{ij}\in[0,1]$ under the corpus's clustering criterion, and these scores provide the numerical basis of the pairwise evidence.
Algorithm~\ref{alg:acquire} in Appendix~\ref{app:algorithms} lists the full procedure, from retrieval to the evidence targets and the relation graph.

A single score, however, does not reveal how many judgments support a pair.
Since the same unordered pair can be judged with either text as the query
text, we collect its observations under the canonical pair $\{i,j\}$ and let
$p_{\{i,j\}}^{(r)}$ denote the $r$th of them. We then retain both the
accumulated score $s_{ij}=\sum_r p_{\{i,j\}}^{(r)}$ and the observation count
$n_{ij}=\#\{r\}$. The accumulated score records the LLM's total support for
the pair, while the count records how many judgments contribute to it.
Together they define the Beta evidence target
\begin{equation}
  a_{ij}=\epsilon+s_{ij},
  \qquad
  b_{ij}=\epsilon+\bigl(n_{ij}-s_{ij}\bigr),
  \label{eq:target}
\end{equation}
with prior weight $\epsilon$. Its mean $a_{ij}/(a_{ij}+b_{ij})$ summarizes the
scores, while the concentration $a_{ij}+b_{ij}=2\epsilon+n_{ij}$
retains the observation count. Pairs with the same mean and different numbers
of observations consequently have different target concentrations, so the
target separates a pair judged once from a pair judged from both sides with
the same average score. These targets supervise the clustering representations
in Section~\ref{sec:method:repr}.

Beyond defining the targets, the same judgments specify where relational
context can propagate. Each text is a node, and the judgment of candidate $j$
for query text $i$ gives a directed relation from $i$ to $j$. Each such
relation receives the acceptance indicator
\begin{equation}
  r_{ij}=\mathbf{1}\!\left[p_{ij}\ge\tau\right],
  \label{eq:thresh}
\end{equation}
where $\tau$ is the graph acceptance threshold, and $j$ becomes an outgoing
neighbour of $i$ when $r_{ij}=1$. The directed edge set
$E^{+}=\{(i,j):r_{ij}=1\}$ retains the query-to-candidate direction; the
reverse judgment, when queried, determines its own edge. Accepted relations
receive unit weight, and each node receives a self-loop,
\begin{align}
  A^{\mathrm{raw}}_{ij}
  &=\mathbf{1}\!\left[(i,j)\in E^{+}\right],
  \label{eq:rawadj}\\
  A^{\mathrm{raw}}_{ii}&=1,
  \label{eq:selfloop}
\end{align}
and message passing uses the row-normalized adjacency
\begin{equation}
  A_{ij}
  =\frac{A^{\mathrm{raw}}_{ij}}{\sum_{k}A^{\mathrm{raw}}_{ik}} .
  \label{eq:rownorm}
\end{equation}
Each row of $A$ averages a node's features with those of its accepted outgoing
neighbours. Thresholding thus selects the relations used for message passing,
whereas all queried pairs, including those below $\tau$, retain their evidence targets.

The relation graph determines where relational context can travel. How far it
travels is learned from the evidence targets in Section~\ref{sec:method:repr}.

\subsection{Evidence-Supervised Representation Learning}
\label{sec:method:repr}

Within the relation graph, each query text is directly connected to the
candidates it accepts at threshold $\tau$. Texts of the same category that
were never retrieved together therefore share no edge, and they are related
through paths of intermediate texts instead. The graph encoder first
transforms the frozen text features with a message-passing network $f_\theta$
over $A$,
\begin{equation}
  H=f_\theta(X,A),
  \label{eq:graphencoder}
\end{equation}
where $H$ holds one hidden vector per text. A propagation of $t$ steps,
$A^tH$, averages the features of texts within $t$ edges, since $A$ includes self-loops. Propagating
farther reaches more texts of the same category, but it also crosses more of the
accepted relations that join different categories, and the best range depends
on the density of the graph and the granularity of the clustering criterion.
The encoder therefore combines the propagation steps with a weight $c_t$ for
each step $t=0,\dots,L$, where $L$ is the maximum number of steps,
\begin{equation}
  U=\sum_{t=0}^{L}c_t\,A^tH,
  \qquad
  z_i=\frac{u_i}{\|u_i\|_2},
  \label{eq:encoder}
\end{equation}
as in generalized PageRank graph filters~\citep{chien2021gprgnn}. The weights
are non-negative and sum to one, and they are obtained from a vector of
trainable parameters $\gamma=(\gamma_0,\dots,\gamma_L)$ by a softmax,
\begin{equation}
  c_t=\frac{\exp(\gamma_t)}{\sum_{s=0}^{L}\exp(\gamma_s)}.
  \label{eq:hopweights}
\end{equation}
They are trained jointly with $f_\theta$ under the evidence targets, so the
range of relational context in $z_i$ is learned for each corpus instead of
being set in advance.

For the evidence to shape these representations, each queried pair needs a
prediction from its two representations that can be compared with its evidence target. A pairwise predictor $g_\psi$ produces this prediction
from the two representations, their absolute difference, and their
elementwise product,
\begin{equation}
  h_{ij}=\left[z_i\Vert z_j\Vert |z_i-z_j|\Vert(z_i\odot z_j)\right],
  \label{eq:pairfeat}
\end{equation}
and produces
\begin{equation}
  (o^+_{ij},o^-_{ij})=g_\psi(h_{ij}).
  \label{eq:pairoutput}
\end{equation}
Since the target carries both a mean and a concentration, the prediction is
likewise a Beta distribution. Following the evidential parametrization in
prior work~\citep{sensoy2018evidential}, the two outputs define
\begin{equation}
  \alpha_{ij}=1+\softp(o^+_{ij}),\qquad
  \beta_{ij}=1+\softp(o^-_{ij}).
  \label{eq:betaparam}
\end{equation}
To align the mean and concentration of this prediction with those of the
target, the parameters $(\theta,\gamma,\psi)$ jointly minimize the KL divergence between the Beta distributions over the canonical queried pairs $\mathcal P$,
\begin{equation}
  \mathcal L=\frac{1}{|\mathcal P|}
  \sum_{\{i,j\}\in\mathcal P}
  \KL\!\left(\Beta(a_{ij},b_{ij})\,\|\,
  \Beta(\alpha_{ij},\beta_{ij})\right).
  \label{eq:kl}
\end{equation}

Every queried pair back-propagates through the same graph encoder, so
the weights $c_t$, which set how far relational context travels, are fitted to
the evidence of all queried pairs. The resulting $z_i$ is the representation delivered to a downstream clusterer,
and Algorithm~\ref{alg:learn} in Appendix~\ref{app:algorithms} summarizes this training procedure.

\subsection{Representation Distillation for Full-Corpus Extension}
\label{sec:method:cluster}

The graph encoder represents the texts of the queried corpus as nodes of the
relation graph, whereas a text that receives no LLM request has no judgments
and hence no node in that graph. Its frozen embedding, however, lies in the
same space as those of the queried texts. The queried texts provide pairs of
frozen embeddings $x_i$ and learned representations $z_i$, and a per-text
mapping fitted on these pairs extends the learned space to new texts without
further LLM requests or graph-encoder training. Following graph-to-MLP
distillation~\citep{zhang2022graphless}, we train a student encoder $q_\phi$
to predict the graph encoder's representation $z_i$ from each queried text's
frozen embedding $x_i$ by mean squared error. The fitted student then maps
an unqueried text with frozen embedding $x_i$ to $\hat z_i = q_\phi(x_i)$ in
the learned space, where a downstream clustering procedure can use it.
Algorithm~\ref{alg:extend} in Appendix~\ref{app:algorithms} gives the distillation, the inference on unqueried
texts and the clustering.

The LLM is thus queried once for the queried corpus, with $bN$ requests, and
the fitted student encodes each new text with a single forward pass.

\section{Experiments}
\label{sec:experiments}

We evaluate the two design choices introduced in Section~\ref{sec:intro}
through clustering quality on the queried corpus and full-corpus extension to
unqueried texts. We then ablate the main components, examine downstream clusterers, and vary the LLM request budget. Further analyses vary graph construction,
judge model, and judge instruction.

\subsection{Setup}
\label{sec:experiments:setup}

\paragraph{Datasets.}
Table~\ref{tab:datasets} lists the seven datasets. BANK77~\citep{casanueva-etal-2020-efficient}, CLINC150~\citep{larson-etal-2019-evaluation}, MASSIVE-I~\citep{fitzgerald-etal-2023-massive} and MTOP-I~\citep{li-etal-2021-mtop} label intents at different granularities; MASSIVE-D relabels the MASSIVE-I utterances with scenarios, MTOP-D groups the MTOP intents into domains, and REDDIT groups posts into communities. The main comparison uses each test corpus, and six datasets also have a larger corpus from the same distribution for full-corpus extension. Every method receives the label count $K$.

\paragraph{Models.}
We use four frozen text embedding models: TF-IDF, all-MiniLM-L6-v2 (MiniLM)~\citep{reimers-gurevych-2019-sentence}, instructor-large (Instructor)~\citep{su-etal-2023-one}, and e5-large-v2 (E5)~\citep{wang2024textembeddingsweaklysupervisedcontrastive}. The LLM judge is Gemma-4-E4B-IT~\citep{gemmateam2026gemma4technicalreport}, and every LLM-guided comparison method queries the same model.

\paragraph{Comparison methods.}
Plain applies K-Means to the frozen embeddings without LLM guidance. The LLM-guided comparisons follow the three families in the Introduction: IDAS~\citep{de-raedt-etal-2023-idas} uses generated text; ClusterLLM~\citep{zhang-etal-2023-clusterllm} and Cequel~\citep{wang-etal-2025-cequel} use pairwise or triplet feedback; SPILL~\citep{lin-etal-2025-spill} and LeBoT~\citep{lin-etal-2026-llms} use LLM-based neighbour selection. ClusterLLM fine-tunes a neural encoder and is therefore evaluated only with neural embeddings. We report clustering accuracy after optimal one-to-one label matching and NMI as percentages, with equal weight for each dataset in the reported means.

\paragraph{Implementation details.}
Unless stated otherwise, \method retrieves $b=6$ blocks of $w=10$ candidates for each text, which requires $6N$ requests, accepts a relation at threshold $\tau=0.4$, and averages results over five random seeds. Architecture, hardware, and the settings of the comparison methods are given in Appendix~\ref{app:repro}.

We first evaluate the clustering representations learned on the queried
corpus.

\begin{table}[t]
\centering
\small
\setlength{\tabcolsep}{4.5pt}
\caption{The seven Datasets. Test is the corpus the main comparison is run on;
the full corpus is the larger same-distribution split used by the extension, and
is absent where the test corpus is already the complete labelled set, as for
REDDIT. In the Test column MASSIVE-I and MASSIVE-D label the same
$2{,}974$ utterances, and MTOP-I and MTOP-D the same $4{,}386$, so each pair
differs only in the granularity of its label space. The full-corpus sizes are
those of the public releases.}
\label{tab:datasets}
\begin{tabular}{llr rr}
\toprule
Dataset & Label space & $K$ & Test & Full corpus \\
\midrule
BANK77 & intent & 77 & 3{,}080 & 9{,}993 \\
CLINC150 & intent & 150 & 4{,}500 & 18{,}350 \\
MASSIVE-I & intent & 59 & 2{,}974 & 13{,}547 \\
MASSIVE-D & scenario & 18 & 2{,}974 & 13{,}547 \\
MTOP-I & intent & 102 & 4{,}386 & 17{,}902 \\
MTOP-D & domain & 11 & 4{,}386 & 17{,}902 \\
REDDIT & community & 50 & 3{,}217 & -- \\
\bottomrule
\end{tabular}
\end{table}

\begin{table*}[!t]
\centering
\footnotesize
\setlength{\tabcolsep}{2.5pt}
\caption{Clustering quality on each evaluation corpus (the Test column of Table~\ref{tab:datasets}). Values are percentages, with standard deviations over seeds in parentheses. Bold marks the best method and underlining the second best.}
\label{tab:main}
\begin{tabular}{llcccccccc}
\toprule
\multirow{2}{*}{Dataset} & \multirow{2}{*}{Method} & \multicolumn{2}{c}{TF-IDF} & \multicolumn{2}{c}{MiniLM} & \multicolumn{2}{c}{Instructor} & \multicolumn{2}{c}{E5} \\
\cmidrule(lr){3-4} \cmidrule(lr){5-6} \cmidrule(lr){7-8} \cmidrule(lr){9-10}
& & ACC & NMI & ACC & NMI & ACC & NMI & ACC & NMI \\
\midrule
\multirow{7}{*}{BANK77} & Plain & 31.93\,{\scriptsize(1.90)} & 53.18\,{\scriptsize(1.78)} & 61.63\,{\scriptsize(1.77)} & 79.26\,{\scriptsize(1.03)} & 51.47\,{\scriptsize(1.30)} & 73.50\,{\scriptsize(0.23)} & 60.57\,{\scriptsize(1.02)} & 77.97\,{\scriptsize(0.20)} \\
 & ClusterLLM & -- & -- & 63.90\,{\scriptsize(1.30)} & 80.70\,{\scriptsize(0.55)} & \underline{68.27\,{\scriptsize(1.88)}} & \underline{83.09\,{\scriptsize(0.33)}} & 67.82\,{\scriptsize(1.28)} & 82.02\,{\scriptsize(0.41)} \\
 & IDAS & 50.26\,{\scriptsize(1.49)} & 71.52\,{\scriptsize(0.16)} & 65.41\,{\scriptsize(2.19)} & 81.55\,{\scriptsize(0.37)} & 61.39\,{\scriptsize(0.70)} & 79.04\,{\scriptsize(0.43)} & 65.82\,{\scriptsize(0.59)} & 81.24\,{\scriptsize(0.23)} \\
 & Cequel & 32.73\,{\scriptsize(0.13)} & 54.74\,{\scriptsize(0.11)} & 58.78\,{\scriptsize(0.20)} & 78.00\,{\scriptsize(0.13)} & 52.77\,{\scriptsize(0.54)} & 74.89\,{\scriptsize(0.24)} & 59.24\,{\scriptsize(0.30)} & 78.22\,{\scriptsize(0.18)} \\
 & SPILL & 52.02\,{\scriptsize(1.12)} & 72.62\,{\scriptsize(0.33)} & 70.60\,{\scriptsize(2.03)} & 84.66\,{\scriptsize(0.39)} & 64.74\,{\scriptsize(0.84)} & 81.20\,{\scriptsize(0.11)} & 70.94\,{\scriptsize(0.89)} & 84.44\,{\scriptsize(0.40)} \\
 & LeBoT & \underline{66.99\,{\scriptsize(1.11)}} & \underline{82.69\,{\scriptsize(0.26)}} & \underline{72.29\,{\scriptsize(0.64)}} & \underline{85.42\,{\scriptsize(0.13)}} & 68.24\,{\scriptsize(0.34)} & 82.85\,{\scriptsize(0.02)} & \underline{71.10\,{\scriptsize(0.48)}} & \underline{85.00\,{\scriptsize(0.10)}} \\
 & Ours & \textbf{71.38\,{\scriptsize(0.72)}} & \textbf{84.90\,{\scriptsize(0.29)}} & \textbf{73.05\,{\scriptsize(1.90)}} & \textbf{86.37\,{\scriptsize(0.22)}} & \textbf{69.14\,{\scriptsize(0.36)}} & \textbf{84.41\,{\scriptsize(0.40)}} & \textbf{73.67\,{\scriptsize(1.07)}} & \textbf{86.62\,{\scriptsize(0.15)}} \\
\midrule
\multirow{7}{*}{CLINC150} & Plain & 48.09\,{\scriptsize(1.27)} & 70.04\,{\scriptsize(0.94)} & 72.70\,{\scriptsize(0.30)} & 89.30\,{\scriptsize(0.13)} & 75.06\,{\scriptsize(1.40)} & 90.36\,{\scriptsize(0.73)} & 75.93\,{\scriptsize(0.71)} & 91.61\,{\scriptsize(0.17)} \\
 & ClusterLLM & -- & -- & 74.08\,{\scriptsize(0.68)} & 89.87\,{\scriptsize(0.24)} & 80.77\,{\scriptsize(0.57)} & 93.30\,{\scriptsize(0.06)} & 80.06\,{\scriptsize(1.47)} & 92.76\,{\scriptsize(0.70)} \\
 & IDAS & 69.99\,{\scriptsize(0.37)} & 85.43\,{\scriptsize(0.16)} & 81.50\,{\scriptsize(1.20)} & 92.52\,{\scriptsize(0.20)} & 80.65\,{\scriptsize(1.82)} & 93.01\,{\scriptsize(0.25)} & 81.16\,{\scriptsize(1.30)} & 93.25\,{\scriptsize(0.21)} \\
 & Cequel & 45.22\,{\scriptsize(0.00)} & 69.24\,{\scriptsize(0.00)} & 71.24\,{\scriptsize(0.00)} & 88.61\,{\scriptsize(0.00)} & 72.76\,{\scriptsize(0.07)} & 89.93\,{\scriptsize(0.03)} & 73.19\,{\scriptsize(0.15)} & 90.82\,{\scriptsize(0.06)} \\
 & SPILL & 68.77\,{\scriptsize(1.62)} & 84.92\,{\scriptsize(0.30)} & 84.27\,{\scriptsize(0.84)} & 93.39\,{\scriptsize(0.34)} & 83.42\,{\scriptsize(0.66)} & 93.73\,{\scriptsize(0.09)} & 84.07\,{\scriptsize(1.71)} & 94.44\,{\scriptsize(0.19)} \\
 & LeBoT & \textbf{86.61\,{\scriptsize(0.62)}} & \underline{93.35\,{\scriptsize(0.09)}} & \textbf{89.23\,{\scriptsize(0.19)}} & \underline{95.38\,{\scriptsize(0.06)}} & \underline{87.75\,{\scriptsize(0.76)}} & \underline{94.72\,{\scriptsize(0.16)}} & \textbf{89.59\,{\scriptsize(0.64)}} & \textbf{95.81\,{\scriptsize(0.02)}} \\
 & Ours & \underline{84.20\,{\scriptsize(1.12)}} & \textbf{93.73\,{\scriptsize(0.12)}} & \underline{87.84\,{\scriptsize(0.66)}} & \textbf{95.59\,{\scriptsize(0.17)}} & \textbf{88.40\,{\scriptsize(0.83)}} & \textbf{95.37\,{\scriptsize(0.05)}} & \underline{88.20\,{\scriptsize(0.68)}} & \underline{95.73\,{\scriptsize(0.17)}} \\
\midrule
\multirow{7}{*}{MASSIVE-D} & Plain & 34.35\,{\scriptsize(2.19)} & 33.53\,{\scriptsize(2.21)} & 57.55\,{\scriptsize(1.76)} & 63.45\,{\scriptsize(0.87)} & 50.76\,{\scriptsize(2.35)} & 59.98\,{\scriptsize(2.08)} & 58.55\,{\scriptsize(8.53)} & 65.10\,{\scriptsize(3.65)} \\
 & ClusterLLM & -- & -- & 59.29\,{\scriptsize(4.15)} & 64.41\,{\scriptsize(1.60)} & 57.74\,{\scriptsize(2.40)} & 67.90\,{\scriptsize(1.01)} & 57.30\,{\scriptsize(2.59)} & 65.30\,{\scriptsize(1.37)} \\
 & IDAS & 47.58\,{\scriptsize(1.28)} & 51.93\,{\scriptsize(1.20)} & 52.43\,{\scriptsize(3.48)} & 61.51\,{\scriptsize(2.30)} & \underline{59.59\,{\scriptsize(3.83)}} & 64.83\,{\scriptsize(0.96)} & 62.35\,{\scriptsize(0.66)} & 64.88\,{\scriptsize(0.30)} \\
 & Cequel & 28.31\,{\scriptsize(0.00)} & 28.36\,{\scriptsize(0.00)} & 56.54\,{\scriptsize(0.04)} & 61.99\,{\scriptsize(0.04)} & 51.00\,{\scriptsize(0.19)} & 59.56\,{\scriptsize(0.25)} & 61.16\,{\scriptsize(2.08)} & 67.71\,{\scriptsize(0.75)} \\
 & SPILL & 46.59\,{\scriptsize(4.68)} & 51.29\,{\scriptsize(0.58)} & \underline{68.20\,{\scriptsize(1.53)}} & \underline{72.29\,{\scriptsize(1.13)}} & 59.29\,{\scriptsize(2.87)} & 65.68\,{\scriptsize(1.15)} & \underline{64.78\,{\scriptsize(2.44)}} & \underline{71.95\,{\scriptsize(1.11)}} \\
 & LeBoT & \underline{59.64\,{\scriptsize(2.78)}} & \underline{64.98\,{\scriptsize(1.94)}} & 66.58\,{\scriptsize(1.28)} & 71.29\,{\scriptsize(0.68)} & 57.68\,{\scriptsize(0.53)} & \underline{68.40\,{\scriptsize(0.18)}} & 61.15\,{\scriptsize(1.34)} & 69.25\,{\scriptsize(0.40)} \\
 & Ours & \textbf{74.17\,{\scriptsize(2.58)}} & \textbf{74.51\,{\scriptsize(1.04)}} & \textbf{72.91\,{\scriptsize(2.97)}} & \textbf{74.70\,{\scriptsize(0.94)}} & \textbf{72.31\,{\scriptsize(1.90)}} & \textbf{75.64\,{\scriptsize(1.07)}} & \textbf{77.87\,{\scriptsize(1.90)}} & \textbf{77.25\,{\scriptsize(0.61)}} \\
\midrule
\multirow{7}{*}{MASSIVE-I} & Plain & 29.16\,{\scriptsize(0.42)} & 47.00\,{\scriptsize(1.31)} & 53.62\,{\scriptsize(0.93)} & 70.60\,{\scriptsize(0.32)} & 51.94\,{\scriptsize(2.22)} & 70.42\,{\scriptsize(0.69)} & 50.83\,{\scriptsize(1.22)} & 70.33\,{\scriptsize(0.42)} \\
 & ClusterLLM & -- & -- & 53.80\,{\scriptsize(2.38)} & 70.74\,{\scriptsize(0.25)} & 55.90\,{\scriptsize(3.02)} & 73.68\,{\scriptsize(0.40)} & 56.12\,{\scriptsize(0.83)} & 74.39\,{\scriptsize(0.46)} \\
 & IDAS & 41.44\,{\scriptsize(1.45)} & 59.35\,{\scriptsize(0.42)} & 58.59\,{\scriptsize(1.68)} & 74.05\,{\scriptsize(0.51)} & 61.01\,{\scriptsize(2.01)} & 75.36\,{\scriptsize(0.35)} & 60.82\,{\scriptsize(1.80)} & 75.95\,{\scriptsize(0.59)} \\
 & Cequel & 26.16\,{\scriptsize(0.00)} & 45.00\,{\scriptsize(0.00)} & 52.87\,{\scriptsize(0.03)} & 69.35\,{\scriptsize(0.06)} & 52.08\,{\scriptsize(0.27)} & 70.39\,{\scriptsize(0.18)} & 50.01\,{\scriptsize(0.45)} & 70.59\,{\scriptsize(0.21)} \\
 & SPILL & 42.05\,{\scriptsize(1.42)} & 62.01\,{\scriptsize(0.47)} & 61.35\,{\scriptsize(2.43)} & 76.17\,{\scriptsize(0.30)} & 57.35\,{\scriptsize(2.31)} & 74.40\,{\scriptsize(0.69)} & 61.97\,{\scriptsize(1.74)} & \underline{77.61\,{\scriptsize(0.59)}} \\
 & LeBoT & \underline{58.07\,{\scriptsize(0.47)}} & \underline{73.87\,{\scriptsize(0.09)}} & \underline{65.87\,{\scriptsize(0.32)}} & \underline{77.21\,{\scriptsize(0.09)}} & \underline{65.13\,{\scriptsize(0.70)}} & \underline{76.69\,{\scriptsize(0.16)}} & \underline{64.76\,{\scriptsize(0.82)}} & 76.81\,{\scriptsize(0.39)} \\
 & Ours & \textbf{65.60\,{\scriptsize(1.21)}} & \textbf{77.12\,{\scriptsize(0.17)}} & \textbf{67.64\,{\scriptsize(1.12)}} & \textbf{78.84\,{\scriptsize(0.27)}} & \textbf{66.89\,{\scriptsize(1.35)}} & \textbf{79.07\,{\scriptsize(0.23)}} & \textbf{66.12\,{\scriptsize(1.08)}} & \textbf{79.04\,{\scriptsize(0.54)}} \\
\midrule
\multirow{7}{*}{MTOP-D} & Plain & 52.95\,{\scriptsize(5.44)} & 51.41\,{\scriptsize(6.90)} & 90.54\,{\scriptsize(0.13)} & 84.53\,{\scriptsize(0.12)} & 89.21\,{\scriptsize(3.99)} & 84.07\,{\scriptsize(2.94)} & 92.51\,{\scriptsize(0.45)} & 87.76\,{\scriptsize(0.30)} \\
 & ClusterLLM & -- & -- & 91.40\,{\scriptsize(0.04)} & 85.55\,{\scriptsize(0.03)} & \underline{91.06\,{\scriptsize(5.04)}} & \textbf{88.37\,{\scriptsize(1.89)}} & 91.35\,{\scriptsize(4.57)} & 87.42\,{\scriptsize(3.29)} \\
 & IDAS & 68.54\,{\scriptsize(1.56)} & 62.74\,{\scriptsize(1.20)} & 84.91\,{\scriptsize(1.34)} & 81.34\,{\scriptsize(0.76)} & 90.10\,{\scriptsize(0.03)} & 86.73\,{\scriptsize(0.09)} & 78.54\,{\scriptsize(1.21)} & 81.02\,{\scriptsize(0.44)} \\
 & Cequel & 51.44\,{\scriptsize(0.00)} & 52.56\,{\scriptsize(0.03)} & 83.00\,{\scriptsize(0.01)} & 81.94\,{\scriptsize(0.02)} & \textbf{93.17\,{\scriptsize(0.08)}} & 86.66\,{\scriptsize(0.10)} & 81.25\,{\scriptsize(0.04)} & 83.08\,{\scriptsize(0.16)} \\
 & SPILL & 65.22\,{\scriptsize(2.87)} & 66.24\,{\scriptsize(1.35)} & \underline{93.08\,{\scriptsize(0.79)}} & \underline{88.63\,{\scriptsize(0.84)}} & 89.48\,{\scriptsize(5.22)} & \underline{87.33\,{\scriptsize(2.85)}} & \underline{92.89\,{\scriptsize(0.02)}} & \underline{88.63\,{\scriptsize(0.05)}} \\
 & LeBoT & \underline{79.79\,{\scriptsize(4.25)}} & \underline{79.30\,{\scriptsize(0.80)}} & 90.17\,{\scriptsize(6.87)} & 88.32\,{\scriptsize(2.94)} & 83.08\,{\scriptsize(3.72)} & 83.27\,{\scriptsize(1.24)} & 90.18\,{\scriptsize(3.65)} & 87.54\,{\scriptsize(2.61)} \\
 & Ours & \textbf{92.40\,{\scriptsize(4.15)}} & \textbf{89.21\,{\scriptsize(1.82)}} & \textbf{94.72\,{\scriptsize(0.20)}} & \textbf{90.50\,{\scriptsize(0.34)}} & 90.71\,{\scriptsize(3.46)} & 87.03\,{\scriptsize(2.11)} & \textbf{93.21\,{\scriptsize(0.79)}} & \textbf{88.87\,{\scriptsize(0.30)}} \\
\midrule
\multirow{7}{*}{MTOP-I} & Plain & 26.40\,{\scriptsize(1.45)} & 57.80\,{\scriptsize(0.95)} & 29.84\,{\scriptsize(1.59)} & 66.93\,{\scriptsize(0.29)} & 31.49\,{\scriptsize(0.64)} & 69.36\,{\scriptsize(0.20)} & 33.34\,{\scriptsize(0.55)} & 70.81\,{\scriptsize(0.36)} \\
 & ClusterLLM & -- & -- & 31.65\,{\scriptsize(0.23)} & 68.52\,{\scriptsize(0.59)} & 33.61\,{\scriptsize(0.57)} & 71.57\,{\scriptsize(0.71)} & 34.70\,{\scriptsize(0.37)} & 74.11\,{\scriptsize(0.59)} \\
 & IDAS & 36.30\,{\scriptsize(0.77)} & 71.02\,{\scriptsize(0.12)} & 35.51\,{\scriptsize(1.14)} & 72.35\,{\scriptsize(0.15)} & 38.78\,{\scriptsize(1.42)} & \underline{74.70\,{\scriptsize(0.17)}} & 38.08\,{\scriptsize(0.71)} & 74.04\,{\scriptsize(0.31)} \\
 & Cequel & 28.81\,{\scriptsize(0.13)} & 57.35\,{\scriptsize(0.15)} & 30.86\,{\scriptsize(0.09)} & 67.60\,{\scriptsize(0.09)} & 31.51\,{\scriptsize(0.25)} & 69.78\,{\scriptsize(0.05)} & 34.09\,{\scriptsize(0.15)} & 70.93\,{\scriptsize(0.05)} \\
 & SPILL & 34.82\,{\scriptsize(0.99)} & 67.22\,{\scriptsize(0.70)} & 37.86\,{\scriptsize(0.77)} & 72.66\,{\scriptsize(0.22)} & 37.44\,{\scriptsize(0.56)} & 73.36\,{\scriptsize(0.39)} & 38.77\,{\scriptsize(1.45)} & 73.91\,{\scriptsize(0.33)} \\
 & LeBoT & \textbf{44.03\,{\scriptsize(1.23)}} & \underline{73.55\,{\scriptsize(0.33)}} & \underline{42.09\,{\scriptsize(1.44)}} & \underline{74.60\,{\scriptsize(0.07)}} & \underline{43.52\,{\scriptsize(0.59)}} & 74.50\,{\scriptsize(0.15)} & \textbf{45.54\,{\scriptsize(0.91)}} & \underline{75.32\,{\scriptsize(0.15)}} \\
 & Ours & \underline{41.96\,{\scriptsize(0.38)}} & \textbf{77.49\,{\scriptsize(0.16)}} & \textbf{45.04\,{\scriptsize(1.76)}} & \textbf{79.11\,{\scriptsize(0.32)}} & \textbf{44.58\,{\scriptsize(1.46)}} & \textbf{78.89\,{\scriptsize(0.23)}} & \underline{43.38\,{\scriptsize(1.51)}} & \textbf{78.80\,{\scriptsize(0.36)}} \\
\midrule
\multirow{7}{*}{REDDIT} & Plain & 11.85\,{\scriptsize(0.41)} & 17.99\,{\scriptsize(0.51)} & 43.18\,{\scriptsize(2.28)} & 50.10\,{\scriptsize(1.08)} & 48.48\,{\scriptsize(0.65)} & 56.27\,{\scriptsize(0.50)} & 44.50\,{\scriptsize(0.47)} & 53.57\,{\scriptsize(0.63)} \\
 & ClusterLLM & -- & -- & 43.54\,{\scriptsize(2.58)} & 50.82\,{\scriptsize(1.02)} & \textbf{55.63\,{\scriptsize(0.75)}} & \textbf{63.49\,{\scriptsize(0.27)}} & 45.50\,{\scriptsize(1.77)} & 54.42\,{\scriptsize(0.35)} \\
 & IDAS & 17.83\,{\scriptsize(0.75)} & 26.72\,{\scriptsize(0.25)} & 30.30\,{\scriptsize(0.38)} & 40.62\,{\scriptsize(0.09)} & 37.84\,{\scriptsize(0.23)} & 47.45\,{\scriptsize(0.08)} & 28.82\,{\scriptsize(0.84)} & 39.70\,{\scriptsize(0.57)} \\
 & Cequel & 13.26\,{\scriptsize(0.06)} & 19.46\,{\scriptsize(0.09)} & 41.29\,{\scriptsize(0.33)} & 49.35\,{\scriptsize(0.33)} & 48.51\,{\scriptsize(0.14)} & 56.36\,{\scriptsize(0.44)} & 41.76\,{\scriptsize(0.63)} & 51.92\,{\scriptsize(0.60)} \\
 & SPILL & 21.32\,{\scriptsize(0.92)} & 28.31\,{\scriptsize(0.16)} & 46.48\,{\scriptsize(1.85)} & 53.56\,{\scriptsize(0.79)} & 46.69\,{\scriptsize(0.30)} & 54.36\,{\scriptsize(0.51)} & 48.97\,{\scriptsize(1.17)} & 56.30\,{\scriptsize(0.36)} \\
 & LeBoT & \underline{40.39\,{\scriptsize(0.84)}} & \underline{45.36\,{\scriptsize(0.28)}} & \underline{52.24\,{\scriptsize(0.33)}} & \underline{56.60\,{\scriptsize(0.18)}} & 52.16\,{\scriptsize(0.09)} & 57.44\,{\scriptsize(0.05)} & \underline{52.92\,{\scriptsize(0.68)}} & \underline{57.90\,{\scriptsize(0.09)}} \\
 & Ours & \textbf{42.42\,{\scriptsize(0.88)}} & \textbf{49.02\,{\scriptsize(0.27)}} & \textbf{53.37\,{\scriptsize(0.83)}} & \textbf{59.07\,{\scriptsize(0.38)}} & \underline{52.30\,{\scriptsize(0.86)}} & \underline{59.02\,{\scriptsize(0.34)}} & \textbf{53.91\,{\scriptsize(0.58)}} & \textbf{60.26\,{\scriptsize(0.38)}} \\
\bottomrule
\end{tabular}
\end{table*}

\subsection{Main Clustering Results}
\label{sec:experiments:main}

The first question is whether the complete framework learns clustering
representations that improve on existing LLM-guided methods on the queried
corpus. Table~\ref{tab:main} shows that \method has the highest mean ACC and
NMI with all four frozen embedders and exceeds the strongest baseline in $46$
of $56$ dataset--embedder--metric cells. The largest aggregate gain occurs with
TF-IDF, whose mean ACC rises from $33.53$ on the frozen embedding to $67.45$.
With E5, \method improves over LeBoT from $67.89$ to $70.91$ ACC and from $78.23$ to $80.94$ NMI. Appendix~\ref{app:mainnotes} relates this gain to LeBoT's design and examines the datasets that share their texts.

We then examine whether the learned clustering space extends beyond the
queried corpus through representation distillation.

\begin{table*}[!t]
\centering
\scriptsize
\setlength{\tabcolsep}{3pt}
\caption{Same-distribution extension to the full corpora in Table~\ref{tab:datasets}. LeBoT and \method are run in the distilled setup of Section~\ref{sec:method:cluster}: their judgments come from a disjoint queried corpus, whose size relative to the corpus clustered here is given by Observed, and a student encoder trained on that queried corpus then encodes every evaluation item, none of which receives an LLM request. The other methods run natively on every evaluation item. Bold marks the best method and underlining the second best.}
\label{tab:extend}
\begin{tabular}{lllcccccccc}
\toprule
\multirow{2}{*}{Dataset} & \multirow{2}{*}{Observed} & \multirow{2}{*}{Method} & \multicolumn{2}{c}{TF-IDF} & \multicolumn{2}{c}{MiniLM} & \multicolumn{2}{c}{Instructor} & \multicolumn{2}{c}{E5} \\
\cmidrule(lr){4-5} \cmidrule(lr){6-7} \cmidrule(lr){8-9} \cmidrule(lr){10-11}
& & & ACC & NMI & ACC & NMI & ACC & NMI & ACC & NMI \\
\midrule
\multirow{7}{*}{BANK77} & \multirow{5}{*}{100.00\%} & Plain & 28.68\,{\scriptsize(1.29)} & 47.56\,{\scriptsize(1.14)} & 61.26\,{\scriptsize(1.26)} & 77.33\,{\scriptsize(0.53)} & 52.21\,{\scriptsize(1.76)} & 71.74\,{\scriptsize(0.82)} & 58.45\,{\scriptsize(1.09)} & 75.86\,{\scriptsize(0.31)} \\
 &  & ClusterLLM & -- & -- & 63.54\,{\scriptsize(0.69)} & 78.33\,{\scriptsize(0.25)} & \textbf{66.05\,{\scriptsize(1.55)}} & \textbf{81.19\,{\scriptsize(0.40)}} & 61.91\,{\scriptsize(1.61)} & 77.56\,{\scriptsize(0.71)} \\
 &  & IDAS & 45.38\,{\scriptsize(1.16)} & 67.39\,{\scriptsize(0.69)} & 61.01\,{\scriptsize(1.14)} & 78.11\,{\scriptsize(0.43)} & 58.96\,{\scriptsize(1.14)} & 76.31\,{\scriptsize(0.35)} & 62.12\,{\scriptsize(1.58)} & 77.59\,{\scriptsize(0.32)} \\
 &  & Cequel & 28.68\,{\scriptsize(0.21)} & 47.59\,{\scriptsize(0.31)} & 60.28\,{\scriptsize(0.00)} & 76.74\,{\scriptsize(0.00)} & 48.13\,{\scriptsize(0.07)} & 70.57\,{\scriptsize(0.04)} & 58.34\,{\scriptsize(0.05)} & 75.86\,{\scriptsize(0.07)} \\
 &  & SPILL & 46.04\,{\scriptsize(0.67)} & 65.92\,{\scriptsize(0.47)} & \underline{68.64\,{\scriptsize(2.20)}} & \textbf{82.72\,{\scriptsize(0.41)}} & 60.94\,{\scriptsize(1.10)} & 78.60\,{\scriptsize(0.52)} & 66.81\,{\scriptsize(0.96)} & \underline{81.43\,{\scriptsize(0.18)}} \\
\addlinespace
 & \multirow{2}{*}{30.82\%} & LeBoT & \underline{55.16\,{\scriptsize(0.25)}} & \underline{72.78\,{\scriptsize(0.05)}} & 67.33\,{\scriptsize(0.25)} & \underline{82.27\,{\scriptsize(0.07)}} & 62.30\,{\scriptsize(0.27)} & 80.03\,{\scriptsize(0.14)} & \underline{67.46\,{\scriptsize(0.34)}} & \textbf{82.65\,{\scriptsize(0.09)}} \\
 &  & Ours & \textbf{61.88\,{\scriptsize(0.60)}} & \textbf{75.95\,{\scriptsize(0.22)}} & \textbf{69.41\,{\scriptsize(0.25)}} & 82.16\,{\scriptsize(0.22)} & \underline{66.01\,{\scriptsize(0.69)}} & \underline{80.71\,{\scriptsize(0.20)}} & \textbf{69.82\,{\scriptsize(0.73)}} & \textbf{82.65\,{\scriptsize(0.29)}} \\
\midrule
\multirow{7}{*}{CLINC150} & \multirow{5}{*}{100.00\%} & Plain & 41.77\,{\scriptsize(1.88)} & 61.26\,{\scriptsize(1.34)} & 76.81\,{\scriptsize(2.08)} & 90.28\,{\scriptsize(0.48)} & 78.51\,{\scriptsize(0.67)} & 91.55\,{\scriptsize(0.23)} & 79.41\,{\scriptsize(0.74)} & 92.76\,{\scriptsize(0.27)} \\
 &  & ClusterLLM & -- & -- & 77.33\,{\scriptsize(1.73)} & 90.41\,{\scriptsize(0.35)} & 82.53\,{\scriptsize(0.82)} & \textbf{93.96\,{\scriptsize(0.28)}} & 83.50\,{\scriptsize(2.59)} & 93.76\,{\scriptsize(0.30)} \\
 &  & IDAS & \underline{69.93\,{\scriptsize(1.13)}} & \textbf{83.93\,{\scriptsize(0.32)}} & \underline{84.00\,{\scriptsize(1.53)}} & \textbf{93.09\,{\scriptsize(0.33)}} & 82.46\,{\scriptsize(1.24)} & 93.46\,{\scriptsize(0.22)} & 84.66\,{\scriptsize(1.01)} & \underline{94.17\,{\scriptsize(0.34)}} \\
 &  & Cequel & 39.43\,{\scriptsize(0.00)} & 59.67\,{\scriptsize(0.00)} & 76.59\,{\scriptsize(0.31)} & 89.97\,{\scriptsize(0.04)} & 78.10\,{\scriptsize(0.02)} & 91.12\,{\scriptsize(0.01)} & 82.44\,{\scriptsize(0.00)} & 93.31\,{\scriptsize(0.00)} \\
 &  & SPILL & 57.60\,{\scriptsize(1.46)} & 76.19\,{\scriptsize(0.82)} & 82.17\,{\scriptsize(0.56)} & \underline{92.89\,{\scriptsize(0.27)}} & \textbf{83.34\,{\scriptsize(0.75)}} & \underline{93.81\,{\scriptsize(0.05)}} & \textbf{85.22\,{\scriptsize(0.76)}} & \textbf{94.53\,{\scriptsize(0.42)}} \\
\addlinespace
 & \multirow{2}{*}{25.00\%} & LeBoT & 58.19\,{\scriptsize(0.54)} & 79.41\,{\scriptsize(0.13)} & 66.33\,{\scriptsize(0.29)} & 87.50\,{\scriptsize(0.06)} & 69.49\,{\scriptsize(0.28)} & 89.28\,{\scriptsize(0.10)} & 69.25\,{\scriptsize(0.31)} & 89.81\,{\scriptsize(0.11)} \\
 &  & Ours & \textbf{72.16\,{\scriptsize(0.94)}} & \underline{83.14\,{\scriptsize(0.28)}} & \textbf{84.98\,{\scriptsize(0.30)}} & 92.11\,{\scriptsize(0.20)} & \underline{83.32\,{\scriptsize(0.83)}} & 92.34\,{\scriptsize(0.24)} & \underline{84.68\,{\scriptsize(0.53)}} & 92.89\,{\scriptsize(0.15)} \\
\midrule
\multirow{7}{*}{MASSIVE-D} & \multirow{5}{*}{100.00\%} & Plain & 31.34\,{\scriptsize(1.78)} & 28.65\,{\scriptsize(1.84)} & 59.20\,{\scriptsize(3.16)} & 64.86\,{\scriptsize(1.19)} & 56.07\,{\scriptsize(2.40)} & 61.82\,{\scriptsize(1.07)} & 60.34\,{\scriptsize(2.53)} & 64.63\,{\scriptsize(2.07)} \\
 &  & ClusterLLM & -- & -- & 56.88\,{\scriptsize(1.42)} & 63.60\,{\scriptsize(0.83)} & 61.07\,{\scriptsize(1.11)} & \underline{68.03\,{\scriptsize(0.22)}} & 60.60\,{\scriptsize(2.53)} & 68.26\,{\scriptsize(0.88)} \\
 &  & IDAS & 42.79\,{\scriptsize(1.06)} & 43.63\,{\scriptsize(0.61)} & 64.02\,{\scriptsize(1.67)} & 64.59\,{\scriptsize(0.76)} & \underline{62.01\,{\scriptsize(2.02)}} & 65.24\,{\scriptsize(0.99)} & \underline{64.95\,{\scriptsize(2.01)}} & 64.80\,{\scriptsize(0.79)} \\
 &  & Cequel & 27.88\,{\scriptsize(0.02)} & 26.48\,{\scriptsize(0.01)} & 59.92\,{\scriptsize(0.04)} & 64.68\,{\scriptsize(0.01)} & 52.62\,{\scriptsize(0.35)} & 58.60\,{\scriptsize(0.51)} & 56.42\,{\scriptsize(0.55)} & 60.66\,{\scriptsize(0.46)} \\
 &  & SPILL & 40.40\,{\scriptsize(2.03)} & 43.37\,{\scriptsize(0.16)} & \underline{66.56\,{\scriptsize(1.25)}} & \underline{71.42\,{\scriptsize(0.25)}} & 58.84\,{\scriptsize(1.36)} & 66.48\,{\scriptsize(1.21)} & 62.49\,{\scriptsize(3.09)} & \underline{69.64\,{\scriptsize(1.66)}} \\
\addlinespace
 & \multirow{2}{*}{22.00\%} & LeBoT & \underline{56.82\,{\scriptsize(0.49)}} & \underline{55.74\,{\scriptsize(0.20)}} & 60.12\,{\scriptsize(1.03)} & 64.98\,{\scriptsize(0.44)} & 54.89\,{\scriptsize(1.21)} & 64.76\,{\scriptsize(0.63)} & 57.38\,{\scriptsize(0.47)} & 65.07\,{\scriptsize(0.27)} \\
 &  & Ours & \textbf{71.61\,{\scriptsize(1.01)}} & \textbf{65.05\,{\scriptsize(0.41)}} & \textbf{75.28\,{\scriptsize(1.85)}} & \textbf{72.64\,{\scriptsize(0.69)}} & \textbf{73.71\,{\scriptsize(1.64)}} & \textbf{72.85\,{\scriptsize(0.72)}} & \textbf{76.35\,{\scriptsize(1.57)}} & \textbf{73.94\,{\scriptsize(0.70)}} \\
\midrule
\multirow{7}{*}{MASSIVE-I} & \multirow{5}{*}{100.00\%} & Plain & 28.69\,{\scriptsize(0.70)} & 42.81\,{\scriptsize(0.91)} & 53.94\,{\scriptsize(0.69)} & 69.51\,{\scriptsize(0.52)} & 52.45\,{\scriptsize(1.12)} & 69.42\,{\scriptsize(0.40)} & 56.52\,{\scriptsize(1.15)} & 71.44\,{\scriptsize(0.24)} \\
 &  & ClusterLLM & -- & -- & 53.89\,{\scriptsize(0.49)} & 69.62\,{\scriptsize(0.13)} & 59.77\,{\scriptsize(1.61)} & 74.13\,{\scriptsize(0.68)} & 57.20\,{\scriptsize(1.32)} & 73.40\,{\scriptsize(0.27)} \\
 &  & IDAS & 44.28\,{\scriptsize(1.29)} & 58.52\,{\scriptsize(0.36)} & 58.47\,{\scriptsize(2.33)} & 73.37\,{\scriptsize(0.25)} & 60.77\,{\scriptsize(0.92)} & 74.45\,{\scriptsize(0.23)} & 57.32\,{\scriptsize(1.38)} & 75.10\,{\scriptsize(0.51)} \\
 &  & Cequel & 25.53\,{\scriptsize(0.00)} & 39.11\,{\scriptsize(0.00)} & 53.90\,{\scriptsize(0.09)} & 69.12\,{\scriptsize(0.10)} & 51.03\,{\scriptsize(0.78)} & 68.94\,{\scriptsize(0.54)} & 54.50\,{\scriptsize(0.08)} & 70.78\,{\scriptsize(0.13)} \\
 &  & SPILL & 37.13\,{\scriptsize(0.10)} & 53.78\,{\scriptsize(0.62)} & 58.66\,{\scriptsize(0.90)} & 73.56\,{\scriptsize(0.20)} & 58.88\,{\scriptsize(1.84)} & 73.50\,{\scriptsize(0.12)} & 61.85\,{\scriptsize(1.83)} & 76.08\,{\scriptsize(0.50)} \\
\addlinespace
 & \multirow{2}{*}{22.01\%} & LeBoT & \underline{54.52\,{\scriptsize(0.23)}} & \underline{63.88\,{\scriptsize(0.09)}} & \underline{64.65\,{\scriptsize(0.34)}} & \underline{73.68\,{\scriptsize(0.13)}} & \underline{63.35\,{\scriptsize(0.41)}} & \textbf{75.28\,{\scriptsize(0.16)}} & \underline{66.19\,{\scriptsize(0.23)}} & \textbf{76.23\,{\scriptsize(0.21)}} \\
 &  & Ours & \textbf{57.64\,{\scriptsize(0.52)}} & \textbf{66.61\,{\scriptsize(0.15)}} & \textbf{66.12\,{\scriptsize(0.63)}} & \textbf{74.63\,{\scriptsize(0.21)}} & \textbf{65.96\,{\scriptsize(0.49)}} & \underline{75.04\,{\scriptsize(0.21)}} & \textbf{66.36\,{\scriptsize(1.29)}} & \underline{76.14\,{\scriptsize(0.13)}} \\
\midrule
\multirow{7}{*}{MTOP-D} & \multirow{5}{*}{100.00\%} & Plain & 52.94\,{\scriptsize(0.60)} & 52.44\,{\scriptsize(3.06)} & 91.02\,{\scriptsize(0.18)} & 85.00\,{\scriptsize(0.10)} & 89.53\,{\scriptsize(5.37)} & 84.57\,{\scriptsize(3.44)} & 91.56\,{\scriptsize(4.12)} & 87.79\,{\scriptsize(1.47)} \\
 &  & ClusterLLM & -- & -- & 92.10\,{\scriptsize(0.02)} & 86.32\,{\scriptsize(0.04)} & \textbf{93.82\,{\scriptsize(0.01)}} & \textbf{89.12\,{\scriptsize(0.01)}} & 90.62\,{\scriptsize(5.65)} & 86.79\,{\scriptsize(3.40)} \\
 &  & IDAS & 62.37\,{\scriptsize(0.83)} & 58.85\,{\scriptsize(1.73)} & 85.68\,{\scriptsize(0.00)} & 78.47\,{\scriptsize(0.00)} & 79.76\,{\scriptsize(4.54)} & 81.14\,{\scriptsize(1.87)} & 73.79\,{\scriptsize(3.85)} & 76.84\,{\scriptsize(1.48)} \\
 &  & Cequel & 46.35\,{\scriptsize(0.01)} & 47.30\,{\scriptsize(0.00)} & 90.94\,{\scriptsize(0.25)} & 84.93\,{\scriptsize(0.13)} & 83.83\,{\scriptsize(0.05)} & 82.16\,{\scriptsize(0.08)} & \underline{91.73\,{\scriptsize(0.06)}} & 86.69\,{\scriptsize(0.11)} \\
 &  & SPILL & 58.62\,{\scriptsize(3.09)} & 58.95\,{\scriptsize(2.12)} & \underline{93.79\,{\scriptsize(0.66)}} & \textbf{89.62\,{\scriptsize(0.31)}} & \underline{92.12\,{\scriptsize(5.76)}} & \underline{88.68\,{\scriptsize(3.37)}} & 91.21\,{\scriptsize(5.56)} & \textbf{88.79\,{\scriptsize(2.11)}} \\
\addlinespace
 & \multirow{2}{*}{24.58\%} & LeBoT & \underline{76.37\,{\scriptsize(1.08)}} & \underline{73.90\,{\scriptsize(0.39)}} & 89.24\,{\scriptsize(0.50)} & 85.39\,{\scriptsize(0.62)} & 85.81\,{\scriptsize(0.39)} & 83.01\,{\scriptsize(0.67)} & 84.44\,{\scriptsize(0.34)} & 84.06\,{\scriptsize(0.30)} \\
 &  & Ours & \textbf{91.37\,{\scriptsize(0.29)}} & \textbf{83.37\,{\scriptsize(0.52)}} & \textbf{93.97\,{\scriptsize(0.18)}} & \underline{88.73\,{\scriptsize(0.24)}} & 91.02\,{\scriptsize(3.18)} & 86.72\,{\scriptsize(1.66)} & \textbf{93.26\,{\scriptsize(0.74)}} & \underline{88.67\,{\scriptsize(0.24)}} \\
\midrule
\multirow{7}{*}{MTOP-I} & \multirow{5}{*}{100.00\%} & Plain & 28.26\,{\scriptsize(1.06)} & 55.95\,{\scriptsize(0.82)} & 32.33\,{\scriptsize(0.72)} & 67.32\,{\scriptsize(0.21)} & 33.19\,{\scriptsize(1.31)} & 69.25\,{\scriptsize(0.48)} & 33.71\,{\scriptsize(0.92)} & 71.48\,{\scriptsize(0.54)} \\
 &  & ClusterLLM & -- & -- & 33.20\,{\scriptsize(1.30)} & 68.58\,{\scriptsize(0.32)} & 33.62\,{\scriptsize(0.81)} & 70.97\,{\scriptsize(0.72)} & 36.48\,{\scriptsize(1.08)} & 74.43\,{\scriptsize(0.41)} \\
 &  & IDAS & 35.99\,{\scriptsize(0.77)} & 68.50\,{\scriptsize(0.16)} & 36.54\,{\scriptsize(0.65)} & 72.41\,{\scriptsize(0.29)} & 37.72\,{\scriptsize(1.01)} & 72.24\,{\scriptsize(0.17)} & 39.45\,{\scriptsize(1.24)} & 74.58\,{\scriptsize(0.35)} \\
 &  & Cequel & 27.83\,{\scriptsize(0.05)} & 55.77\,{\scriptsize(0.01)} & 31.55\,{\scriptsize(0.10)} & 66.80\,{\scriptsize(0.04)} & 32.55\,{\scriptsize(0.05)} & 68.34\,{\scriptsize(0.02)} & 35.61\,{\scriptsize(0.07)} & 71.82\,{\scriptsize(0.05)} \\
 &  & SPILL & 33.62\,{\scriptsize(1.73)} & 63.26\,{\scriptsize(0.34)} & 37.08\,{\scriptsize(1.69)} & 70.98\,{\scriptsize(0.34)} & 37.69\,{\scriptsize(2.79)} & 72.30\,{\scriptsize(0.53)} & 38.39\,{\scriptsize(0.85)} & 73.17\,{\scriptsize(0.45)} \\
\addlinespace
 & \multirow{2}{*}{24.58\%} & LeBoT & \textbf{46.57\,{\scriptsize(0.35)}} & \underline{69.90\,{\scriptsize(0.10)}} & \underline{43.77\,{\scriptsize(0.59)}} & \underline{73.15\,{\scriptsize(0.12)}} & \textbf{46.29\,{\scriptsize(0.39)}} & \underline{75.28\,{\scriptsize(0.12)}} & \textbf{47.43\,{\scriptsize(0.15)}} & \underline{75.68\,{\scriptsize(0.05)}} \\
 &  & Ours & \underline{39.78\,{\scriptsize(0.88)}} & \textbf{71.71\,{\scriptsize(0.21)}} & \textbf{44.07\,{\scriptsize(0.39)}} & \textbf{75.98\,{\scriptsize(0.15)}} & \underline{45.30\,{\scriptsize(1.54)}} & \textbf{76.91\,{\scriptsize(0.40)}} & \underline{43.33\,{\scriptsize(1.51)}} & \textbf{77.07\,{\scriptsize(0.24)}} \\
\bottomrule
\end{tabular}
\end{table*}

\subsection{Full-Corpus Extension}
\label{sec:experiments:full}

The second question is whether the learned clustering space extends to
unqueried same-distribution texts. Table~\ref{tab:extend} reports results on
six larger corpora. For \method and LeBoT, a student encoder is trained on a
disjoint queried corpus and applied to evaluation texts that receive no LLM
request: LeBoT's request volume of about $11N$ makes a direct pass over corpora
of this size impractical, so it is run in the same distilled setup as \method.
The queried corpora are 22--31\% as large as the de-duplicated
evaluation corpora; the other baselines run on every evaluation item.

The \method student achieves the highest mean on all four embedders. With E5,
it exceeds SPILL by $4.6$ ACC points and the LeBoT student on five of six
datasets, including gains of $19.0$ points on MASSIVE-D and $15.4$ on
CLINC150. MTOP-I is the ACC exception, where the LeBoT student leads by up to
$6.8$ points; the \method student has the highest NMI there with all four
embedders. Across datasets, the largest ACC gap between the queried corpus and the distilled full corpus is $5.1$ points for the three neural embedders and
$12.0$ for TF-IDF. On MASSIVE-D with E5, the student reaches $76.35$ ACC,
compared with $77.87$ on the queried corpus.

Across the four embedders, \method requires 6.9--9.9 seconds of end-to-end runtime
after the LLM responses are available and is the second-fastest method in every
setting, following LeBoT at 6.1--8.6 seconds. Figure~\ref{fig:time} in
Appendix~\ref{app:complexity} reports the runtime of every method, together with
the cost of each stage of the pipeline.

\subsection{Ablation Study}
\label{sec:experiments:ablation}

We ablate the three components that the Introduction attributes the gains to. (1)~\textbf{No graph} replaces $A$ by the identity matrix, so each text is encoded without relational context while the evidence targets are unchanged. (2)~\textbf{Fixed propagation range} sets $c_2=1$ and $c_t=0$ for $t\neq2$, so the representation is computed from $A^2H$ alone and the range of relational context is set in advance instead of learned. (3)~\textbf{Discrete target} queries the LLM again and takes its same-class or different-class output for each pair, so the supervision is a binary decision that carries no evidence. Each variant is paired with the full method on the same dataset, embedding and seed, which gives $140$ paired runs.

\begin{figure}[t]
\centering
\includegraphics[width=\linewidth]{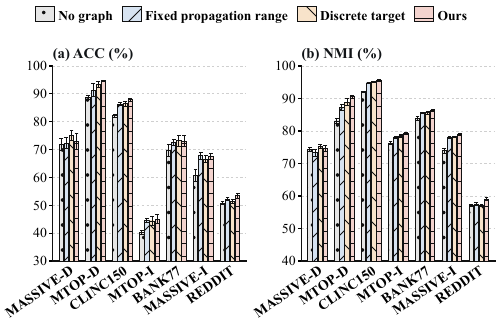}
\Description{Two bar charts side by side compare the full method and the three ablated variants under MiniLM across seven datasets: accuracy on the left, NMI on the right. Error bars show variation across five encoder seeds.}
\caption{Ablation on MiniLM. Bars show the mean performance for each dataset and variant; error bars show the standard deviation across five encoder seeds.}
\label{fig:ablation}
\end{figure}

Figure~\ref{fig:ablation} shows the dataset-wise means and seed variation on MiniLM, and Appendix~\ref{app:ablation} (Figures~\ref{fig:ablation:tfidf}--\ref{fig:ablation:e5}) reports the same comparison for the other three frozen embedders. Without the graph, ACC drops in $26$ of the $28$ cells and NMI in all $28$, by $3.54$ and $2.94$ points on average. With the propagation range fixed, ACC drops in $23$ cells and NMI in $27$, by $1.22$ and $1.29$ points. With the discrete target, ACC drops in $22$ cells and NMI in $27$, by $0.84$ and $0.89$ points. The paired Wilcoxon signed-rank test over the $140$ runs is significant for all three variants on both metrics ($p<10^{-5}$).

\subsection{Downstream Clustering on Learned Representations}
\label{sec:experiments:clusterers}

To examine how the learned space is used downstream, we apply K-Means, Ward
Hierarchical Clustering, and Spectral Clustering to the same MiniLM
representations. The learned representations work with all three procedures:
averaged over the seven datasets, Ward Hierarchical Clustering reaches the highest
mean ($72.38$ ACC and $81.18$ NMI, against $70.66$ and $80.59$ for K-Means).
Appendix~\ref{app:clusterers} reports the dataset-wise results.

\subsection{Query Efficiency}
\label{sec:experiments:budget}

\begin{figure}[t]
\centering
\includegraphics[width=\linewidth]{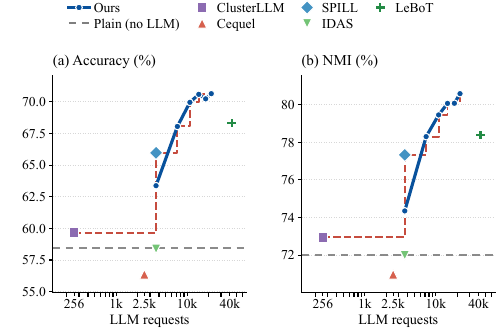}
\Description{Two panels side by side show mean accuracy and NMI against logarithmic LLM
request count for MiniLM, including the \method budget curve, comparison-method
operating points, the no-LLM reference and the empirical Pareto envelope.}
\caption{Request--performance comparison with MiniLM. \method is evaluated from
$1N$ to $6N$ requests; markers show comparison methods at their evaluated
budgets, and the red staircase shows the empirical Pareto envelope. Values are
means over seven datasets.}
\label{fig:budget:minilm}
\end{figure}

Figure~\ref{fig:budget:minilm} compares mean ACC and NMI against the number of
LLM requests for MiniLM; Appendix~\ref{app:budget} (Figures~\ref{fig:budget:tfidf}--\ref{fig:budget:e5}) reports the same comparison
for the other three frozen embedders. \method is evaluated
at block budgets $b=1,\ldots,6$, corresponding to $1N$--$6N$ requests. Each
comparison method appears at its evaluated operating point, while Plain is shown
as a no-LLM reference. For methods whose request budget scales with $N$, the
horizontal coordinate uses the mean corpus size of $3{,}645$ texts across the
seven datasets. The red staircase is the empirical Pareto envelope of the
plotted operating points.

\method reaches the empirical Pareto frontier within $2N$--$3N$ requests ($1N$ under TF-IDF). At $6N$, or $21.9$k requests on average, it has the highest mean ACC and NMI under all four embedders, whereas LeBoT uses $42.5$k requests on average and scores lower in every panel. Appendix~\ref{app:budget} gives the frontier for each embedder and the gain from $1N$ to $6N$.

\subsection{Further Analysis}
\label{sec:experiments:further}

We vary three settings with MiniLM as the frozen encoder; each analysis is given in full in the appendix. Over the acceptance threshold $\tau\in[0.1,0.8]$ the cross-dataset mean stays inside a band of $1.7$ ACC and $0.7$ NMI points, so the adopted $\tau=0.4$ lies on a plateau (Appendix~\ref{app:tau}). Across four judge models on BANK77 and REDDIT, mean ACC ranges from $60.22$ for Gemma-4-E2B-IT to $64.07$ for DeepSeek-V4-Flash, and Gemma-4-E2B-IT has an NMI $3.4$--$4.7$ points below the other three (Appendix~\ref{app:judges}). Adding the corpus setting and a one-sentence statement of what a class corresponds to (V2) raises the seven-dataset mean over the task-only V1 from $67.13$ to $70.65$ ACC and from $78.00$ to $80.60$ NMI, whereas the more explicit V3 falls $1.52$ and $1.15$ points below V2 (Appendix~\ref{app:prompts}).

\section{Conclusion}
\label{sec:conclusion}

\method turns one pass of local LLM judgments into a corpus-level clustering
representation. Modeling each judgment as pairwise evidence gives every queried
pair a target that records both the accumulated score and the number of
observations behind it, and the relations accepted at the threshold form a graph
over the corpus. On this graph the encoder learns, under the same evidence, how
far relational context propagates, so the supervision shapes the representation
itself rather than only the partition. Paired ablations confirm that the graph,
the learned propagation range and the evidence target each contribute. Because
the request count grows as $bN$ and a single pass serves all of training, the
block budget sets the cost directly: the budget sweep places \method on the
empirical Pareto frontier from $2N$ requests under the neural embedders, and a
distilled student extends the learned space to unqueried texts with a single
forward pass and no further LLM requests. Across seven datasets and four frozen
text embeddings, \method obtains the highest mean accuracy and NMI.

\begin{acks}
\zk{This work was supported by the National Natural
Science Foundation of China (62673389) and the National
Social Science Fund of China (23BTJ053).}
\end{acks}

\bibliographystyle{ACM-Reference-Format}
\bibliography{reference}

@inproceedings{tian-etal-2023-just,
  author    = {Tian, Katherine and Mitchell, Eric and Zhou, Allan and Sharma, Archit and Rafailov, Rafael and Yao, Huaxiu and Finn, Chelsea and Manning, Christopher},
  title     = {Just Ask for Calibration: Strategies for Eliciting Calibrated Confidence Scores from Language Models Fine-Tuned with Human Feedback},
  booktitle = {Proceedings of the Conference on Empirical Methods in Natural Language Processing (EMNLP)},
  pages     = {5433--5442},
  year      = {2023}
}

@inproceedings{sensoy2018evidential,
  author    = {Sensoy, Murat and Kaplan, Lance and Kandemir, Melih},
  title     = {Evidential Deep Learning to Quantify Classification Uncertainty},
  booktitle = {Advances in Neural Information Processing Systems (NeurIPS), Vol. 31},
  year      = {2018}
}

@inproceedings{chien2021gprgnn,
  author    = {Chien, Eli and Peng, Jianhao and Li, Pan and Milenkovic, Olgica},
  title     = {Adaptive Universal Generalized {PageRank} Graph Neural Network},
  booktitle = {International Conference on Learning Representations (ICLR)},
  year      = {2021}
}

@inproceedings{zhang2022graphless,
  author    = {Zhang, Shichang and Liu, Yozen and Sun, Yizhou and Shah, Neil},
  title     = {Graph-less Neural Networks: Teaching Old {MLP}s New Tricks via Distillation},
  booktitle = {International Conference on Learning Representations (ICLR)},
  year      = {2022}
}

@inproceedings{de-raedt-etal-2023-idas,
  author    = {De Raedt, Maarten and Godin, Fr{\'e}deric and Demeester, Thomas and Develder, Chris},
  title     = {{IDAS}: Intent Discovery with Abstractive Summarization},
  booktitle = {Proceedings of the Workshop on NLP for Conversational AI (NLP4ConvAI)},
  pages     = {71--88},
  year      = {2023}
}

@inproceedings{sac2026,
  author    = {Diaz-Rodriguez, Jairo},
  title     = {Summaries as Centroids for Interpretable and Scalable Text Clustering},
  booktitle = {International Conference on Learning Representations (ICLR)},
  year      = {2026}
}

@inproceedings{zhang-etal-2023-clusterllm,
  author    = {Zhang, Yuwei and Wang, Zihan and Shang, Jingbo},
  title     = {{C}luster{LLM}: Large Language Models as a Guide for Text Clustering},
  booktitle = {Proceedings of the Conference on Empirical Methods in Natural Language Processing (EMNLP)},
  pages     = {13903--13920},
  year      = {2023}
}

@inproceedings{wang-etal-2025-cequel,
  author    = {Wang, Hongtao and Zhang, Taiyan and Yang, Renchi and Xu, Jianliang},
  title     = {{C}equel: Cost-Effective Querying of Large Language Models for Text Clustering},
  booktitle = {Proceedings of the ACM International Conference on Information and Knowledge Management (CIKM)},
  pages     = {2998--3008},
  year      = {2025}
}

@inproceedings{lin-etal-2026-llms,
  author    = {Lin, I-Fan and Hasibi, Faegheh and Verberne, Suzan},
  title     = {{LLM}s Enable Bag-of-Texts Representations for Short-Text Clustering},
  booktitle = {Proceedings of the Annual Meeting of the Association for Computational Linguistics (ACL)},
  pages     = {6432--6447},
  year      = {2026}
}

@inproceedings{lin-etal-2025-spill,
  author    = {Lin, I-Fan and Hasibi, Faegheh and Verberne, Suzan},
  title     = {{SPILL}: Domain-Adaptive Intent Clustering based on Selection and Pooling with Large Language Models},
  booktitle = {Findings of the Association for Computational Linguistics: ACL},
  pages     = {15723--15737},
  year      = {2025}
}

@inproceedings{li-etal-2025-co,
  author    = {Li, Zetong and Su, Qinliang and Huang, Minhua and Yang, Yin},
  title     = {Co-Evolving {LLM}s and Embedding Models via Density-Guided Preference Optimization for Text Clustering},
  booktitle = {Proceedings of the Conference on Empirical Methods in Natural Language Processing (EMNLP)},
  pages     = {4796--4808},
  year      = {2025}
}

@inproceedings{feng-etal-2024-llmedgerefine,
  author    = {Feng, Zijin and Lin, Luyang and Wang, Lingzhi and Cheng, Hong and Wong, Kam-Fai},
  title     = {{LLME}dge{R}efine: Enhancing Text Clustering with {LLM}-Based Boundary Point Refinement},
  booktitle = {Proceedings of the Conference on Empirical Methods in Natural Language Processing (EMNLP)},
  pages     = {18455--18462},
  year      = {2024}
}

@inproceedings{wagstaff2001cop,
  author    = {Wagstaff, Kiri and Cardie, Claire and Rogers, Seth and Schr{\"o}dl, Stefan},
  title     = {Constrained K-means Clustering with Background Knowledge},
  booktitle = {Proceedings of the International Conference on Machine Learning (ICML)},
  pages     = {577--584},
  year      = {2001}
}

@inproceedings{basu2004active,
  author    = {Basu, Sugato and Banerjee, Arindam and Mooney, Raymond J.},
  title     = {Active Semi-Supervision for Pairwise Constrained Clustering},
  booktitle = {Proceedings of the SIAM International Conference on Data Mining (SDM)},
  pages     = {333--344},
  year      = {2004}
}

@inproceedings{zhang2021deepaligned,
  author    = {Zhang, Hanlei and Xu, Hua and Lin, Ting-En and Lyu, Rui},
  title     = {Discovering New Intents with Deep Aligned Clustering},
  booktitle = {Proceedings of the AAAI Conference on Artificial Intelligence (AAAI)},
  volume    = {35},
  number    = {16},
  pages     = {14365--14373},
  year      = {2021}
}

@inproceedings{zhang-etal-2022-new,
  author    = {Zhang, Yuwei and Zhang, Haode and Zhan, Li-Ming and Wu, Xiao-Ming and Lam, Albert Y. S.},
  title     = {New Intent Discovery with Pre-training and Contrastive Learning},
  booktitle = {Proceedings of the Annual Meeting of the Association for Computational Linguistics (ACL)},
  pages     = {256--269},
  year      = {2022}
}

@inproceedings{yin-wang-2014-dirichlet,
  author    = {Yin, Jianhua and Wang, Jianyong},
  title     = {A {D}irichlet Multinomial Mixture Model-Based Approach for Short Text Clustering},
  booktitle = {Proceedings of the ACM SIGKDD Conference on Knowledge Discovery and Data Mining (KDD)},
  pages     = {233--242},
  year      = {2014}
}

@inproceedings{zhang-etal-2021-supporting,
  author    = {Zhang, Dejiao and Nan, Feng and Wei, Xiaokai and Li, Shang-Wen and Zhu, Henghui and McKeown, Kathleen and Nallapati, Ramesh and Arnold, Andrew O. and Xiang, Bing},
  title     = {Supporting Clustering with Contrastive Learning},
  booktitle = {Proceedings of the Conference of the North American Chapter of the Association for Computational Linguistics (NAACL)},
  pages     = {5419--5430},
  year      = {2021}
}

@article{xiong-etal-2014-active,
  author    = {Xiong, Sicheng and Azimi, Javad and Fern, Xiaoli Z.},
  title     = {Active Learning of Constraints for Semi-Supervised Clustering},
  journal   = {IEEE Transactions on Knowledge and Data Engineering},
  volume    = {26},
  number    = {1},
  pages     = {43--54},
  year      = {2014}
}

@inproceedings{zhang-etal-2025-semi-supervised,
  author    = {Zhang, Chao and Xu, Deng and Chen, Chunlin and Li, Huaxiong},
  title     = {Semi-supervised Multi-view Clustering with Active Constraints},
  booktitle = {Proceedings of the ACM SIGKDD Conference on Knowledge Discovery and Data Mining (KDD)},
  pages     = {1903--1912},
  year      = {2025}
}

@inproceedings{vonluxburg-etal-2012-clustering,
  author    = {von Luxburg, Ulrike and Williamson, Robert C. and Guyon, Isabelle},
  title     = {Clustering: Science or Art?},
  booktitle = {Proceedings of the ICML Workshop on Unsupervised and Transfer Learning, PMLR, Vol. 27},
  pages     = {65--80},
  year      = {2012}
}

@inproceedings{fitzgerald-etal-2023-massive,
  author    = {FitzGerald, Jack and Hench, Christopher and Peris, Charith and Mackie, Scott and Rottmann, Kay and Sanchez, Ana and Nash, Aaron and Urbach, Liam and Kakarala, Vishesh and Singh, Richa and Ranganath, Swetha and Crist, Laurie and Britan, Misha and Leeuwis, Wouter and Tur, Gokhan and Natarajan, Prem},
  title     = {{MASSIVE}: A 1{M}-Example Multilingual Natural Language Understanding Dataset with 51 Typologically-Diverse Languages},
  booktitle = {Proceedings of the Annual Meeting of the Association for Computational Linguistics (ACL)},
  pages     = {4277--4302},
  year      = {2023}
}

@inproceedings{muennighoff-etal-2023-mteb,
  author    = {Muennighoff, Niklas and Tazi, Nouamane and Magne, Loic and Reimers, Nils},
  title     = {{MTEB}: Massive Text Embedding Benchmark},
  booktitle = {Proceedings of the Conference of the European Chapter of the Association for Computational Linguistics (EACL)},
  pages     = {2014--2037},
  year      = {2023}
}

@inproceedings{jia-etal-2026-optimized,
  author    = {Jia, Chaoqi and Wu, Weihong and Guo, Longkun and Lu, Zhigang and Chen, Chao and Ong, Kok-Leong},
  title     = {Optimized Algorithms for Text Clustering with LLM-Generated Constraints},
  booktitle = {Proceedings of the AAAI Conference on Artificial Intelligence (AAAI)},
  pages     = {22229--22237},
  year      = {2026}
}

@inproceedings{wang-etal-2026-nilc,
  author    = {Wang, Hongtao and Yang, Renchi and Lin, Wenqing},
  title     = {{NILC}: Discovering New Intents with {LLM}-assisted Clustering},
  booktitle = {Proceedings of the ACM International Conference on Web Search and Data Mining (WSDM)},
  pages     = {671--680},
  year      = {2026}
}

@inproceedings{cheng-etal-2026-nonparametric,
  author    = {Cheng, Enhao and Zheng, Xiaolong and Li, Jintong and Hou, Juncheng and Liu, Fan and Song, Xuemeng and Gan, Tian and Yin, Jianhua},
  title     = {A Non-parametric Multi-view Model for Short Text Clustering},
  booktitle = {Proceedings of the ACM SIGKDD Conference on Knowledge Discovery and Data Mining (KDD)},
  pages     = {532--543},
  year      = {2026}
}

@inproceedings{casanueva-etal-2020-efficient,
  author    = {Casanueva, I{\~n}igo and Tem{\v{c}}inas, Tadas and Gerz, Daniela and Henderson, Matthew and Vuli{\'c}, Ivan},
  title     = {Efficient Intent Detection with Dual Sentence Encoders},
  booktitle = {Proceedings of the Workshop on NLP for Conversational AI (NLP4ConvAI)},
  pages     = {38--45},
  year      = {2020}
}

@inproceedings{larson-etal-2019-evaluation,
  author    = {Larson, Stefan and Mahendran, Anish and Peper, Joseph J. and Clarke, Christopher and Lee, Andrew and Hill, Parker and Kummerfeld, Jonathan K. and Leach, Kevin and Laurenzano, Michael A. and Tang, Lingjia and Mars, Jason},
  title     = {An Evaluation Dataset for Intent Classification and Out-of-Scope Prediction},
  booktitle = {Proceedings of the Conference on Empirical Methods in Natural Language Processing and the International Joint Conference on Natural Language Processing (EMNLP-IJCNLP)},
  pages     = {1311--1316},
  year      = {2019}
}

@inproceedings{li-etal-2021-mtop,
  author    = {Li, Haoran and Arora, Abhinav and Chen, Shuohui and Gupta, Anchit and Gupta, Sonal and Mehdad, Yashar},
  title     = {{MTOP}: A Comprehensive Multilingual Task-Oriented Semantic Parsing Benchmark},
  booktitle = {Proceedings of the Conference of the European Chapter of the Association for Computational Linguistics (EACL)},
  pages     = {2950--2962},
  year      = {2021}
}

@inproceedings{su-etal-2023-one,
  author    = {Su, Hongjin and Shi, Weijia and Kasai, Jungo and Wang, Yizhong and Hu, Yushi and Ostendorf, Mari and Yih, Wen-tau and Smith, Noah A. and Zettlemoyer, Luke and Yu, Tao},
  title     = {One Embedder, Any Task: Instruction-Finetuned Text Embeddings},
  booktitle = {Findings of the Association for Computational Linguistics: ACL},
  pages     = {1102--1121},
  year      = {2023}
}

@inproceedings{reimers-gurevych-2019-sentence,
  author    = {Reimers, Nils and Gurevych, Iryna},
  title     = {Sentence-{BERT}: Sentence Embeddings Using {S}iamese {BERT}-Networks},
  booktitle = {Proceedings of the Conference on Empirical Methods in Natural Language Processing and the International Joint Conference on Natural Language Processing (EMNLP-IJCNLP)},
  pages     = {3982--3992},
  year      = {2019}
}

@article{wang2024textembeddingsweaklysupervisedcontrastive,
  author    = {Wang, Liang and Yang, Nan and Huang, Xiaolong and Jiao, Binxing and Yang, Linjun and Jiang, Daxin and Majumder, Rangan and Wei, Furu},
  title     = {Text Embeddings by Weakly-Supervised Contrastive Pre-training},
  journal   = {arXiv preprint arXiv:2212.03533},
  year      = {2024}
}

@article{gemmateam2026gemma4technicalreport,
  author    = {{Gemma Team}},
  title     = {Gemma 4 Technical Report},
  journal   = {arXiv preprint arXiv:2607.02770},
  year      = {2026}
}

@inproceedings{yan-etal-2013-biterm,
  author    = {Yan, Xiaohui and Guo, Jiafeng and Lan, Yanyan and Cheng, Xueqi},
  title     = {A Biterm Topic Model for Short Texts},
  booktitle = {Proceedings of the ACM Web Conference (WWW)},
  pages     = {1445--1456},
  year      = {2013}
}

@inproceedings{chen-etal-2022-intent,
  author    = {Chen, Minhua and Jayakumar, Badrinath and Johnston, Michael and Mahmoodi, S. Eman and Pressel, Daniel},
  title     = {Intent Discovery for Enterprise Virtual Assistants: Applications of Utterance Embedding and Clustering to Intent Mining},
  booktitle = {Proceedings of the Conference of the North American Chapter of the Association for Computational Linguistics: Industry Track (NAACL)},
  pages     = {197--208},
  year      = {2022}
}

@inproceedings{kumar-etal-2022-intent,
  author    = {Kumar, Rajat and Patidar, Mayur and Varshney, Vaibhav and Vig, Lovekesh and Shroff, Gautam},
  title     = {Intent Detection and Discovery from User Logs via Deep Semi-Supervised Contrastive Clustering},
  booktitle = {Proceedings of the Conference of the North American Chapter of the Association for Computational Linguistics (NAACL)},
  pages     = {1836--1853},
  year      = {2022}
}

@inproceedings{wang-davidson-2010-active,
  author    = {Wang, Xiang and Davidson, Ian},
  title     = {Active Spectral Clustering},
  booktitle = {Proceedings of the IEEE International Conference on Data Mining (ICDM)},
  pages     = {561--568},
  year      = {2010}
}

@article{zhang-etal-2021-framework,
  author    = {Zhang, Hongjing and Zhan, Tianyang and Basu, Sugato and Davidson, Ian},
  title     = {A Framework for Deep Constrained Clustering},
  journal   = {Data Mining and Knowledge Discovery},
  volume    = {35},
  number    = {2},
  pages     = {593--620},
  year      = {2021}
}

@article{ren-etal-2019-semi,
  author    = {Ren, Yazhou and Hu, Kangrong and Dai, Xinyi and Pan, Lili and Hoi, Steven C. H. and Xu, Zenglin},
  title     = {Semi-Supervised Deep Embedded Clustering},
  journal   = {Neurocomputing},
  volume    = {325},
  pages     = {121--130},
  year      = {2019}
}

@inproceedings{lin-etal-2020-discovering,
  author    = {Lin, Ting-En and Xu, Hua and Zhang, Hanlei},
  title     = {Discovering New Intents via Constrained Deep Adaptive Clustering with Cluster Refinement},
  booktitle = {Proceedings of the AAAI Conference on Artificial Intelligence (AAAI)},
  pages     = {8360--8367},
  year      = {2020}
}

@article{zhang-etal-2024-clustering,
  author    = {Zhang, Hanlei and Xu, Hua and Wang, Xin and Long, Fei and Gao, Kai},
  title     = {A Clustering Framework for Unsupervised and Semi-Supervised New Intent Discovery},
  journal   = {IEEE Transactions on Knowledge and Data Engineering},
  volume    = {36},
  number    = {11},
  pages     = {5468--5481},
  year      = {2024}
}

@inproceedings{hong-etal-2025-dial,
  author    = {Hong, Mengze and Ng, Wailing and Zhang, Chen Jason and Song, Yuanfeng and Jiang, Di},
  title     = {Dial-In {LLM}: Human-Aligned {LLM}-in-the-Loop Intent Clustering for Customer Service Dialogues},
  booktitle = {Proceedings of the Conference on Empirical Methods in Natural Language Processing (EMNLP)},
  pages     = {5885--5900},
  year      = {2025}
}

@inproceedings{islam-2026-reasoning,
  author    = {Islam, Tunazzina},
  title     = {Reasoning-Based Refinement of Unsupervised Text Clusters with {LLM}s},
  booktitle = {Findings of the Association for Computational Linguistics: ACL},
  pages     = {9899--9917},
  year      = {2026}
}

@inproceedings{fan-etal-2024-lanid,
  author    = {Fan, Lu and Pu, Jiashu and Zhang, Rongsheng and Wu, Xiao-Ming},
  title     = {{LANID}: {LLM}-assisted New Intent Discovery},
  booktitle = {Proceedings of the Joint International Conference on Computational Linguistics, Language Resources and Evaluation (LREC-COLING)},
  pages     = {10110--10116},
  year      = {2024}
}

@inproceedings{yang-etal-2024-tec,
  author    = {Yang, Chen and Cao, Bin and Fan, Jing},
  title     = {{TeC}: A Novel Method for Text Clustering with Large Language Models Guidance and Weakly-Supervised Contrastive Learning},
  booktitle = {Proceedings of the International AAAI Conference on Web and Social Media (ICWSM)},
  pages     = {1702--1712},
  year      = {2024}
}

@inproceedings{an-etal-2024-loop,
  author    = {An, Wenbin and Shi, Wenkai and Tian, Feng and Lin, Haonan and Wang, QianYing and Wu, Yaqiang and Cai, Mingxiang and Wang, Luyan and Chen, Yan and Zhu, Haiping and Chen, Ping},
  title     = {Generalized Category Discovery with Large Language Models in the Loop},
  booktitle = {Findings of the Association for Computational Linguistics: ACL},
  pages     = {8653--8665},
  year      = {2024}
}

@inproceedings{fan-etal-2025-ligd,
  author    = {Fan, Kangjia and Zhao, Yilong and Li, Daifeng and Lin, Changze and Zhang, Weijun and Zhong, Zhiwen},
  title     = {{LLM}-Enhanced Generalized Category Discovery via Iterative Graph Diffusion},
  booktitle = {Proceedings of the ACM International Conference on Information and Knowledge Management (CIKM)},
  pages     = {645--654},
  year      = {2025}
}

@article{viswanathan-etal-2024-large,
  author    = {Viswanathan, Vijay and Gashteovski, Kiril and Lawrence, Carolin and Wu, Tongshuang and Neubig, Graham},
  title     = {Large Language Models Enable Few-Shot Clustering},
  journal   = {Transactions of the Association for Computational Linguistics},
  volume    = {12},
  pages     = {321--333},
  year      = {2024}
}

\clearpage

\appendix

\section{Algorithms}
\label{app:algorithms}

Algorithms~\ref{alg:acquire}--\ref{alg:extend} give the pseudocode of the three stages described in Section~\ref{sec:method}: relation acquisition, representation learning on the relation graph, and the extension to unqueried texts.

\begin{algorithm}[t]
\small
\caption{\method{} relation acquisition: from LLM pair judgments to a scored
graph}
\label{alg:acquire}
\KwIn{queried texts $V=\{1,\dots,N\}$; clustering criterion in natural
language; frozen embedder; LLM judge; block budget $b$ and block size $w$;
acceptance threshold $\tau$; prior weight $\epsilon$}
\KwOut{row-normalized relation graph $A$; evidence targets $\{(a_{ij},b_{ij})\}$}
\BlankLine
$x_i\gets$ frozen embedding of text $i$ for all $i\in V$;\quad
$X\gets[x_1,\dots,x_N]^\top$\;
\BlankLine
\ForEach{query text $i\in V$}{
  rank the other texts of $V$ by cosine similarity to $x_i$ and keep the top
  $bw$ candidates\;
  divide them by rank into $b$ disjoint blocks of size $w$\;
  \ForEach{block}{
    send one batched LLM request with the raw texts of $i$ and of the block's
    candidates under the criterion\;
    \ForEach{candidate $j$ in the block}{
      record the score $p_{ij}\in[0,1]$ returned by the judge\;
    }
  }
}
\BlankLine
\ForEach{canonical pair $\{i,j\}$ with at least one recorded score}{
  $s_{ij}\gets\sum_r p^{(r)}_{\{i,j\}}$;\quad $n_{ij}\gets\#\{r\}$\;
  $a_{ij}\gets\epsilon+s_{ij}$;\quad $b_{ij}\gets\epsilon+(n_{ij}-s_{ij})$\;
}
\BlankLine
$E^{+}\gets\{(i,j):p_{ij}\ge\tau\}$\;
$A^{\mathrm{raw}}_{ij}\gets\mathbf{1}[(i,j)\in E^{+}]$,\quad
$A^{\mathrm{raw}}_{ii}\gets1$,\quad
$A_{ij}\gets A^{\mathrm{raw}}_{ij}/\sum_{k}A^{\mathrm{raw}}_{ik}$\;
\end{algorithm}

\begin{algorithm}[t]
\small
\caption{\method{} representation learning on the relation graph}
\label{alg:learn}
\KwIn{relation graph $A$; frozen embeddings $X$; evidence targets $\{(a_{ij},b_{ij})\}$; maximum number of hops $L$}
\KwOut{clustering representations $Z=[z_1,\dots,z_N]^\top$}
\BlankLine
initialize encoder parameters $\theta$, hop logits $\gamma\in\mathbb{R}^{L+1}$ and predictor parameters $\psi$\;
\Repeat{convergence}{
  $H\gets f_\theta(X,A)$\;
  $c_t\gets$ weight of hop $t$ computed from $\gamma$ by Eq.~\eqref{eq:hopweights}, for $t=0,\dots,L$\;
  $U\gets\sum_{t=0}^{L}c_tA^tH$\;
  $z_i\gets u_i/\|u_i\|_2$ for every text $i$\;
  \ForEach{canonical queried pair $\{i,j\}$}{
    $(\alpha_{ij},\beta_{ij})\gets$ Beta parameters that $g_\psi$ predicts from $z_i$ and $z_j$, Eqs.~\eqref{eq:pairfeat}--\eqref{eq:betaparam}\;
  }
  $\mathcal L\gets$ mean over the queried pairs of
  $\KL\bigl(\Beta(a_{ij},b_{ij})\,\|\,\Beta(\alpha_{ij},\beta_{ij})\bigr)$\;
  update $\theta,\gamma,\psi$ to reduce $\mathcal L$\;
}
\end{algorithm}

\begin{algorithm}[t]
\small
\caption{\method{} extension to unqueried texts and clustering}
\label{alg:extend}
\KwIn{representations $Z$; frozen embeddings $X$; unqueried texts $V'$;
cluster count $K$}
\KwOut{student $q_\phi$; cluster assignments of $V$ and of $V'$}
\BlankLine
train $q_\phi$ on the queried rows $(x_i,z_i)$, $i\in V$, to minimize
$\frac{1}{N}\sum_{i\in V}\|q_\phi(x_i)-z_i\|_2^2$ (mean squared error)\;
\BlankLine
\ForEach{text $i\in V'$}{
  embed $i$ with the frozen embedder to obtain $x_i$ and compute $\hat z_i=q_\phi(x_i)$\;
}
\BlankLine
\textbf{queried corpus:} K-Means with $K$ clusters on the rows of $Z$\;
\textbf{full corpus:} K-Means with $K$ clusters on $\hat z_i$, $i\in V'$\;
\end{algorithm}

\section{Sensitivity to the Acceptance Threshold}
\label{app:tau}

\begin{figure}[t]
\centering
\begin{minipage}[t]{0.49\linewidth}
  \centering
  \includegraphics[width=\linewidth]{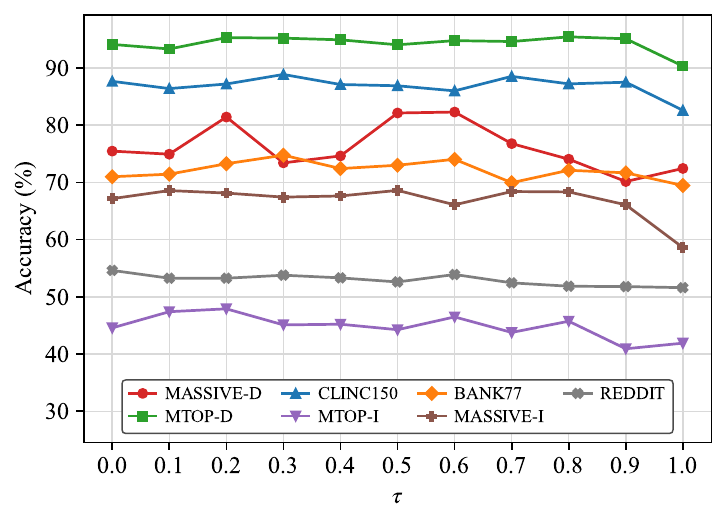}\\[-2pt]
  (a)~Accuracy
\end{minipage}\hfill
\begin{minipage}[t]{0.49\linewidth}
  \centering
  \includegraphics[width=\linewidth]{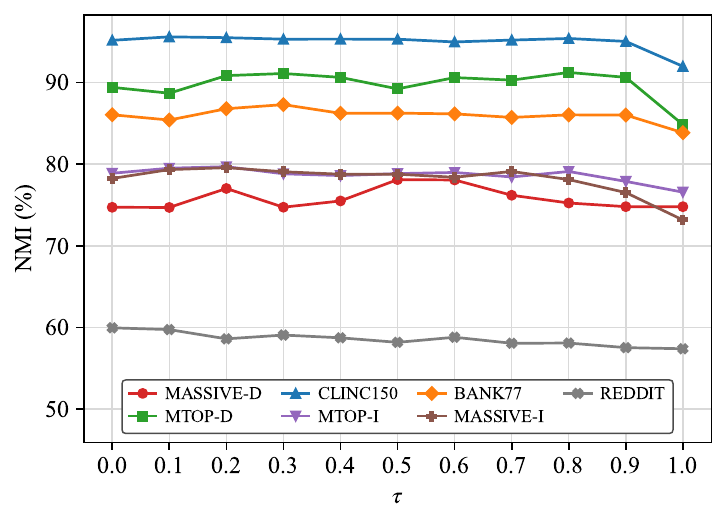}\\[-2pt]
  (b)~NMI
\end{minipage}
\Description{Two panels of line plots, accuracy and NMI against the acceptance
threshold, each showing a flat plateau per dataset that drops as the threshold
approaches one.}
\caption{Threshold sensitivity on MiniLM. Each point retrains the
graph at that threshold. The cross-dataset mean is stable for
$\tau\in[0.1,0.8]$ and declines as $\tau$ approaches 1.}
\label{fig:tau}
\end{figure}

We sweep the acceptance threshold $\tau$ from $0$ to $1$ on MiniLM (Figure~\ref{fig:tau}) to test how graph density affects clustering quality, changing only which edges survive the threshold. Over
$\tau\in[0.1,0.8]$ the cross-dataset mean stays inside a band of $1.7$ points of
accuracy and $0.7$ of NMI, and its two endpoints differ by less than $0.1$, so
the adopted $\tau=0.4$ lies on a plateau. The plateau holds to $\tau=0.8$ and
declines as $\tau$ approaches $1$. At $\tau=1$, accuracy falls by $4.0$ points
and NMI by $3.0$; the mean number of accepted relations per node falls from
$68$ to $0.5$, and between $62\%$ and $99.5\%$ of nodes per dataset have no
accepted relations ($84\%$ on average). Most nodes then retain only their
self-loop, leaving little relational context for propagation.

MASSIVE-D varies most within the plateau, by 8.9 accuracy points between $\tau=0.1$ and $\tau=0.8$; the other datasets vary by 4.8 points or less.

\section{Judge Instruction Versions}
\label{app:prompts}

The three instruction versions of Section~\ref{sec:experiments:further} are instantiated per corpus, replacing the setting and the statement of what a class corresponds to with those of the corpus at hand. We show the three instructions for BANK77.

\subsection{Instructions}

The shared closing rule, which every version carries, is shown once, after the
three versions. Below we give the corpus-specific part of each version.

\begin{promptbox}{V1: task only}
\promptlabel{Instruction}
\prompttext{You rate whether two short texts belong to the same class as the target.}
\end{promptbox}

\begin{promptbox}{V2: task, setting and what a class corresponds to}
\promptlabel{Instruction}
\prompttext{You rate whether two short texts belong to the same class as the
target. They are customer messages sent to a bank's support team. A class is the
specific customer issue that the team would resolve in one handling step.}
\end{promptbox}

\begin{promptbox}{V3: V2 plus granularity and criteria}
\promptlabel{Instruction}
\prompttext{You rate whether two short texts belong to the same class as the
target. They are customer messages sent to a bank's support team. A class is the
specific customer issue that the team would resolve in one handling step. This
taxonomy is fine-grained. Messages that would need different handling are
different classes, even when they share a broad topic. Wording, formality, and
phrasing are irrelevant: the same issue said differently is still the same
class.}
\end{promptbox}

\begin{promptbox}{Shared closing rule (all versions)}
\promptlabel{Output}
\prompttext{For every candidate output p\_same, your probability in [0,1] that
it belongs to the same class as the target under the rule above. Use the full
range: near 1.0 when clearly the same, near 0.0 when clearly different, and
intermediate values when genuinely torn -{}- do not collapse uncertain cases to
0. Rate every candidate ID exactly once. Use only the displayed text. Return
JSON only.}
\end{promptbox}

The user message is identical in all three versions. It names the target, lists
the candidates of one block with their identifiers, and closes with an
instruction to emit one score per identifier:

\begin{promptbox}{User message (BANK77, first candidates of one block)}
\promptlabel{TARGET UTTERANCE:}
\prompttext{My card has not arrived yet.}

\promptlabel{CANDIDATE UTTERANCES (preserve these IDs):}
\prompttext{Candidate c\_2551: My card hasn't arrived yet.\\
Candidate c\_0421: I am waiting for my card to arrive.\\
Candidate c\_1230: i have not received my card\\
\ldots}
\prompttext{Output one p\_same for every candidate ID listed above.}
\end{promptbox}

\subsection{Performance}

\begin{table*}[!t]
\centering
\footnotesize
\setlength{\tabcolsep}{5pt}
\caption{Effect of the judge instruction on clustering quality, MiniLM encoder, averaged over the seven datasets. Numbers are percentages; the parenthesised value is one standard deviation over five seeds. V1 states the task only, V2 adds the corpus setting and a single sentence on what a class corresponds to, and V3 adds granularity and positive and negative criteria. V2 is the configuration of Table~\ref{tab:main}, whose MiniLM column is repeated here. Bold marks the best version for each dataset and metric.}
\label{tab:promptver}
\begin{tabular}{lcccccc}
\toprule
 & \multicolumn{3}{c}{ACC} & \multicolumn{3}{c}{NMI} \\
\cmidrule(lr){2-4}\cmidrule(lr){5-7}
Dataset & V1 & V2 & V3 & V1 & V2 & V3 \\
\midrule
BANK77 & 70.75\,{\scriptsize(0.99)} & \textbf{73.05\,{\scriptsize(1.90)}} & 72.32\,{\scriptsize(1.07)} & 84.88\,{\scriptsize(0.31)} & \textbf{86.37\,{\scriptsize(0.22)}} & 86.05\,{\scriptsize(0.23)} \\
CLINC150 & 86.60\,{\scriptsize(1.26)} & 87.84\,{\scriptsize(0.66)} & \textbf{89.76\,{\scriptsize(0.23)}} & 95.50\,{\scriptsize(0.22)} & 95.59\,{\scriptsize(0.17)} & \textbf{96.35\,{\scriptsize(0.15)}} \\
MASSIVE-I & 64.94\,{\scriptsize(1.91)} & \textbf{67.64\,{\scriptsize(1.12)}} & 62.29\,{\scriptsize(3.26)} & 77.92\,{\scriptsize(0.67)} & \textbf{78.84\,{\scriptsize(0.27)}} & 74.81\,{\scriptsize(1.15)} \\
MASSIVE-D & 63.34\,{\scriptsize(3.93)} & \textbf{72.91\,{\scriptsize(2.97)}} & 67.32\,{\scriptsize(1.87)} & 69.10\,{\scriptsize(1.13)} & \textbf{74.70\,{\scriptsize(0.94)}} & 70.89\,{\scriptsize(1.14)} \\
MTOP-I & 43.07\,{\scriptsize(1.54)} & \textbf{45.04\,{\scriptsize(1.76)}} & 44.16\,{\scriptsize(1.87)} & 77.20\,{\scriptsize(0.70)} & \textbf{79.11\,{\scriptsize(0.32)}} & 78.41\,{\scriptsize(0.53)} \\
MTOP-D & 89.55\,{\scriptsize(2.58)} & 94.72\,{\scriptsize(0.20)} & \textbf{95.03\,{\scriptsize(0.29)}} & 84.21\,{\scriptsize(1.82)} & 90.50\,{\scriptsize(0.34)} & \textbf{90.93\,{\scriptsize(0.42)}} \\
REDDIT & 51.64\,{\scriptsize(0.70)} & \textbf{53.37\,{\scriptsize(0.83)}} & 53.03\,{\scriptsize(1.27)} & 57.19\,{\scriptsize(0.38)} & \textbf{59.07\,{\scriptsize(0.38)}} & 58.71\,{\scriptsize(0.50)} \\
\bottomrule
\end{tabular}
\end{table*}

Table~\ref{tab:promptver} gives the full breakdown behind
Section~\ref{sec:experiments:further}: accuracy and NMI for each dataset and version,
and the standard deviation over five seeds. The comparison
uses MiniLM, the encoder of the threshold analysis in
Appendix~\ref{app:tau}, and the V2 column is the MiniLM column of
Table~\ref{tab:main}. Supplying the corpus setting and the statement of what a
class corresponds to improves all seven datasets on both metrics. V3 leads on only two of them,
CLINC150 and MTOP-D, by less than two points, while V2 leads on the other five
and by more than five points on MASSIVE-D and MASSIVE-I.

The judge instruction fixes how much of the clustering criterion the model is told. We compare three versions that share the request layout of \S\ref{sec:method:acq} and differ only in the system instruction: V1 states the task only, V2 adds the corpus setting and, in a single sentence, what a class corresponds to, and V3 additionally fixes the granularity and states positive and negative criteria. Each version is instantiated per corpus.

On MiniLM, supplying the corpus setting and that single sentence raises the average over the seven datasets from $67.13$ to $70.65$ accuracy and from $78.00$ to $80.60$ NMI, and it improves every dataset on both metrics. The explicit criteria of V3 move the average back down: V3 is $1.52$ accuracy points and $1.15$ NMI points below V2, and it leads on only two of the seven datasets. V2 gives the highest mean on both metrics, and its per-corpus statement of what a class corresponds to gives the judge prior information comparable to that supplied in prior work \citep{wang-etal-2025-cequel}.

\section{Judge Model Size}
\label{app:judges}

To examine how judge choice relates to clustering performance, we run the full
pipeline with four judges on BANK77 and REDDIT, using MiniLM as the frozen
encoder. The judges are Gemma-4-E2B-IT, Gemma-4-E4B-IT, Gemma-4-12B-IT, and
DeepSeek-V4-Flash. Table~\ref{tab:params} reports their configurations; the
set spans dense models of different sizes and a mixture-of-experts model. Every
judge answers the same request set under the same prompt and response schema.

%
\begin{table*}[!t]
\centering
\footnotesize
\setlength{\tabcolsep}{4pt}
\caption{Configurations of the four judges compared in this appendix. The last
column counts the parameters that take part in producing one token.
\textsuperscript{a}\,Total includes the embedding matrix.
\textsuperscript{b}\,Effective count under which the model family is released,
written E for Effective; the model is dense.
\textsuperscript{c}\,Dense model, so every parameter takes part in each token.
\textsuperscript{d}\,Mixture of experts; the expert capacity routed to each
token, against 284B stored.}
\label{tab:params}
\begin{tabular}{llrr}
\toprule
Judge & Architecture & Total & Per token \\
\midrule
Gemma-4-E2B-IT & Dense & 5.1B\textsuperscript{a} & 2.3B\textsuperscript{b} \\
Gemma-4-E4B-IT & Dense & 8B\textsuperscript{a} & 4.5B\textsuperscript{b} \\
Gemma-4-12B-IT & Dense & 11.95B & 11.95B\textsuperscript{c} \\
DeepSeek-V4-Flash & MoE & 284B & 13B\textsuperscript{d} \\
\bottomrule
\end{tabular}
\end{table*}

Table~\ref{tab:judges} reports the clustering quality each judge produces.

\begin{table*}[!t]
\centering
\footnotesize
\setlength{\tabcolsep}{5pt}
\caption{Clustering quality under four judges of increasing size, with MiniLM as the frozen encoder. Values are percentages, with one standard deviation over five seeds in parentheses; the last two columns average the two datasets for each judge. Bold marks the best judge per column.}
\label{tab:judges}
\begin{tabular}{lcccccc}
\toprule
 & \multicolumn{2}{c}{BANK77} & \multicolumn{2}{c}{REDDIT} & \multicolumn{2}{c}{Average} \\
\cmidrule(lr){2-3}\cmidrule(lr){4-5}\cmidrule(lr){6-7}
Judge & ACC & NMI & ACC & NMI & ACC & NMI \\
\midrule
DeepSeek-V4-Flash & \textbf{73.89\,{\scriptsize(1.22)}} & \textbf{86.55\,{\scriptsize(0.19)}} & 54.25\,{\scriptsize(0.49)} & 60.00\,{\scriptsize(0.40)} & \textbf{64.07\,{\scriptsize(0.41)}} & 73.28\,{\scriptsize(0.16)} \\
Gemma-4-12B-IT & 72.13\,{\scriptsize(2.19)} & 86.11\,{\scriptsize(0.54)} & \textbf{55.39\,{\scriptsize(0.56)}} & \textbf{62.01\,{\scriptsize(0.54)}} & 63.76\,{\scriptsize(1.17)} & \textbf{74.06\,{\scriptsize(0.36)}} \\
Gemma-4-E4B-IT & 73.05\,{\scriptsize(2.13)} & 86.37\,{\scriptsize(0.24)} & 53.37\,{\scriptsize(0.93)} & 59.07\,{\scriptsize(0.42)} & 63.21\,{\scriptsize(1.50)} & 72.72\,{\scriptsize(0.29)} \\
Gemma-4-E2B-IT & 71.06\,{\scriptsize(1.42)} & 84.85\,{\scriptsize(0.28)} & 49.38\,{\scriptsize(1.35)} & 53.79\,{\scriptsize(0.78)} & 60.22\,{\scriptsize(1.10)} & 69.32\,{\scriptsize(0.47)} \\
\bottomrule
\end{tabular}
\end{table*}

Mean ACC ranges from $60.22$ for Gemma-4-E2B-IT to $64.07$ for
DeepSeek-V4-Flash, with Gemma-4-E4B-IT at $63.21$ and Gemma-4-12B-IT at
$63.76$. For mean NMI, Gemma-4-E2B-IT is lowest at $69.32$; the other three
judges fall within $1.34$ points of one another, with Gemma-4-12B-IT highest at
$74.06$. Dataset-wise, DeepSeek-V4-Flash leads both metrics on BANK77, while
Gemma-4-12B-IT leads both on REDDIT.

\section{Reproducibility Details}
\label{app:repro}

The code release contains the exact prompts, response schema, preprocessing,
random seeds and configuration files. We summarise the choices needed to
interpret the reported comparisons.

\paragraph{Judging.}
Gemma-4-E4B-IT returns one $p_{ij}\in[0,1]$ per candidate under a
strict JSON schema at temperature $0$. Candidates are presented in blocks of
$w=10$; missing or malformed ratings are treated as unobserved.

The criterion given to the judge is the V2 instruction of
Appendix~\ref{app:prompts} (one sentence of setting and one of what a class
corresponds to), instantiated per corpus: customer messages to a bank's
support team, and the issue the team would resolve in one handling step
(BANK77); requests to a virtual assistant, and the one thing it is asked to do
(CLINC150); spoken commands to a voice assistant, and one instruction it
executes (MASSIVE-I); commands to a voice assistant, and the broad service they
call on (MASSIVE-D); one action applied to one kind of thing (MTOP-I); the broad
area of user need (MTOP-D); and post titles together with the topic-community
they were submitted under (REDDIT). The judge sees the texts and the criterion;
no label is shown to it.

\paragraph{Baselines.}
All comparison methods are run with the same five random seeds as \method, and each follows the implementation in its released repository (Table~\ref{tab:repos}). Plain is Lloyd K-Means with $k$-means++ initialisation on the frozen embedding.
IDAS and SPILL issue one request per item, Cequel at most $2{,}500$ requests per
dataset, ClusterLLM at most $256$, and LeBoT about $11N$; \method uses $6N$ at
its main operating point.
For full-corpus transfer, LeBoT's student predicts a 64-dimensional
principal-component projection of its medoid block.

\begin{table*}[!t]
\centering
\small
\caption{Repositories from which the comparison methods were reproduced. Plain
is Lloyd K-Means and uses no external release.}
\label{tab:repos}
\begin{tabular}{ll}
\toprule
Method & Released implementation \\
\midrule
IDAS & \url{https://github.com/maarten-deraedt/IDAS-intent-discovery-with-abstract-summarization} \\
ClusterLLM & \url{https://github.com/zhang-yu-wei/ClusterLLM} \\
Cequel & \url{https://github.com/HKBU-LAGAS/Cequel} \\
SPILL & \url{https://github.com/tom192180/SPILL_Clustering_LLM} \\
LeBoT & \url{https://github.com/tom192180/BoT_vector} \\
\bottomrule
\end{tabular}
\end{table*}

\paragraph{Implementation.}
The graph encoder has two message-passing layers with output dimension $64$ and propagates over at most $L=4$ steps; the judge runs at temperature $0$, and all experiments ran on a single NVIDIA RTX 5060 Ti. The graph encoder uses hidden width $128$. The pairwise predictor is a two-layer
MLP over \eqref{eq:pairfeat}, and the student encoder has two hidden layers of
$512$ units. Training and clustering ran with Python 3.11 and PyTorch 2.12, and
the judge was served on an NVIDIA A100 40GB. The release provides the complete
environment specification.

\section{Additional Notes on the Main Results}
\label{app:mainnotes}

Like \method, LeBoT begins with candidates retrieved near each target, but it updates its representation through prespecified neighbour pooling, whereas \method learns propagation on the relation graph built from the LLM judgments, with its range learned under pairwise evidence supervision.

On the datasets that share their texts, the E5 margin over the best baseline changes from $+1.4$ to $+13.1$ points between MASSIVE-I ($K{=}59$) and MASSIVE-D ($K{=}18$), and from $-2.2$ to $+0.3$ between MTOP-I ($K{=}102$) and MTOP-D ($K{=}11$).

\section{Downstream Clustering Procedures}
\label{app:clusterers}

\begin{figure*}[t]
\centering
\begin{minipage}[t]{0.49\textwidth}
  \centering
  \includegraphics[width=\linewidth]{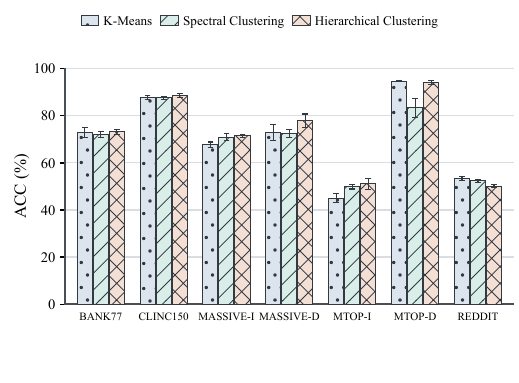}\\[-2pt]
  (a)~ACC
\end{minipage}\hfill
\begin{minipage}[t]{0.49\textwidth}
  \centering
  \includegraphics[width=\linewidth]{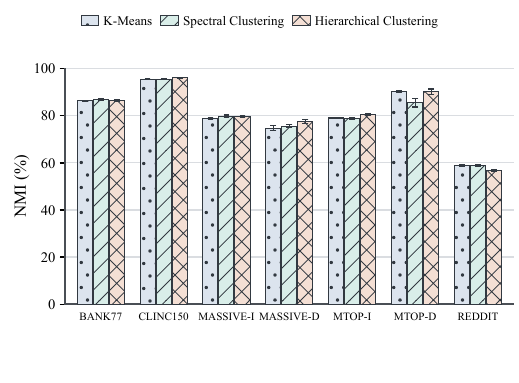}\\[-2pt]
  (b)~NMI
\end{minipage}
\Description{Two bar charts compare K-Means, Spectral Clustering, and Ward
Hierarchical Clustering on the same MiniLM representations across seven
datasets. Error bars show variation across five encoder seeds.}
\caption{Downstream clustering on the same learned MiniLM representations.
Bars show the mean performance for each dataset and clustering procedure;
error bars show the standard deviation across five encoder seeds.}
\label{fig:clusterers}
\end{figure*}

Figure~\ref{fig:clusterers} compares K-Means, Spectral Clustering and Ward Hierarchical Clustering on the same learned MiniLM representations. The comparison covers seven datasets and five encoder seeds; only the clustering procedure changes for each representation. Averaged over the datasets, Ward Hierarchical Clustering reaches $72.38$ ACC and $81.18$ NMI, compared with $70.66$ and $80.59$ for K-Means and $69.85$ and $80.23$ for Spectral Clustering.

\section{Time Complexity}
\label{app:complexity}

Let $N$ be the size of the queried corpus, $b$ the block budget, $w$ the block
size, $d_x$ the embedding dimension, $h$ the encoder width, $L$ the maximum
propagation step and $T$ the number of training epochs.

\paragraph{Judging pass.}
A pass issues $bN$ requests and records $bwN$ scores, from which the evidence
targets and the sparse relation graph are assembled in $O(bwN)$. Ranking the
candidates of each text by cosine similarity in the frozen embedding space
costs $O(N^2d_x)$ once, before the first request.

\paragraph{Training.}
One epoch runs the encoder over the graph and the pairwise predictor over the
$|\mathcal P|=O(bwN)$ queried pairs. With the accepted relations
$\mathrm{nnz}=O(bwN)$ and $L+1$ propagation steps, the graph encoder costs
$O(L\cdot\mathrm{nnz})$ in the relations plus the projection of the $N$ frozen
embeddings, and the predictor costs $O(|\mathcal P|)$. Both counts are linear in
$N$ for a fixed budget, so an epoch is $O(N)$ and training is $O(TN)$.

\paragraph{Extension and clustering.}
The student encoder is trained on the $N$ queried texts in $O(T'Nd_xh)$ for
$T'$ epochs and encodes each unqueried text in a single forward pass of
$O(d_xh)$. Lloyd K-Means over the learned representations adds $O(NKh_zI)$ for
$I$ iterations.

The request count $bN$ is the cost the block budget sets directly, and every
other stage is linear in the corpus size for a fixed budget.

\begin{figure*}[t]
\centering
\includegraphics[width=\textwidth]{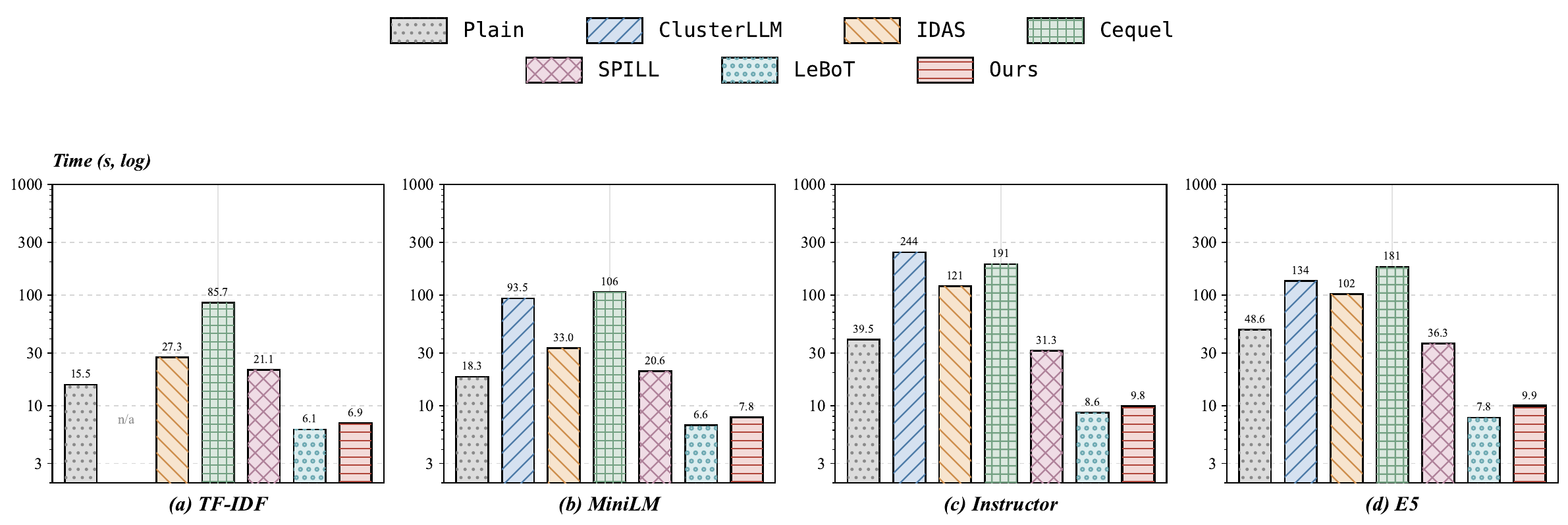}
\Description{Four bar charts compare end-to-end runtime for seven clustering
methods under TF-IDF, MiniLM, Instructor, and E5 embeddings. The vertical axis
is logarithmic.}
\caption{End-to-end runtime excluding LLM request time. LeBoT and \method use
the representation-distillation setup described in
Section~\ref{sec:method:cluster}. The vertical axis uses a logarithmic scale.}
\label{fig:time}
\end{figure*}

\section{Ablation on the Other Embeddings}
\label{app:ablation}

\begin{figure*}[t]
\centering
\includegraphics[width=\textwidth]{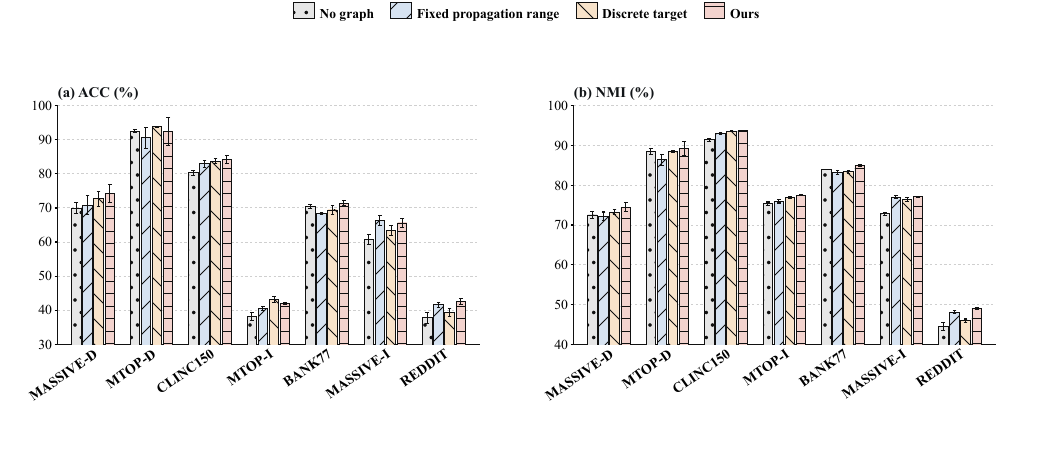}
\Description{Two bar charts side by side compare the full method and the three ablated variants under TF-IDF across seven datasets: accuracy on the left, NMI on the right. Error bars show variation across five encoder seeds.}
\caption{Ablation on TF-IDF. Bars show the mean performance for each dataset and variant; error bars show the standard deviation across five encoder seeds.}
\label{fig:ablation:tfidf}
\end{figure*}

\begin{figure*}[t]
\centering
\includegraphics[width=\textwidth]{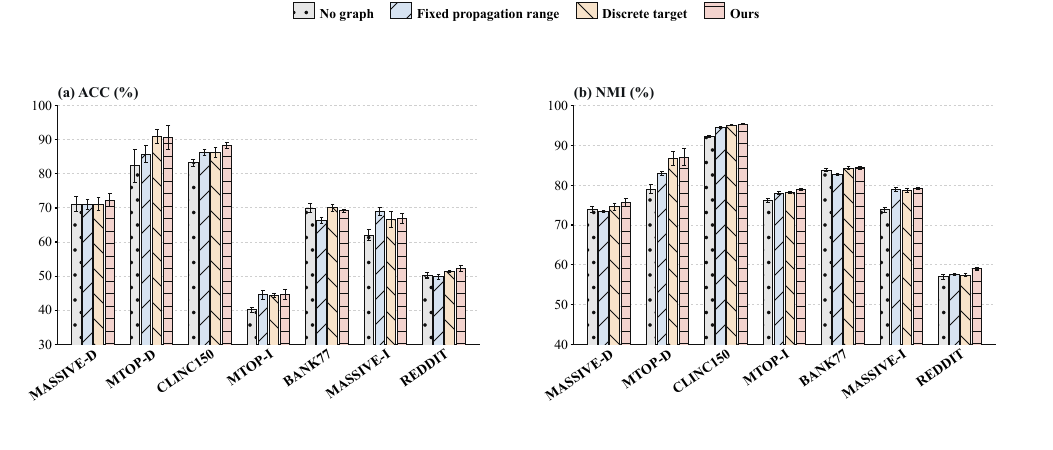}
\Description{Two bar charts side by side compare the full method and the three ablated variants under Instructor across seven datasets: accuracy on the left, NMI on the right. Error bars show variation across five encoder seeds.}
\caption{Ablation on Instructor. Bars show the mean performance for each dataset and variant; error bars show the standard deviation across five encoder seeds.}
\label{fig:ablation:instructor}
\end{figure*}

\begin{figure*}[t]
\centering
\includegraphics[width=\textwidth]{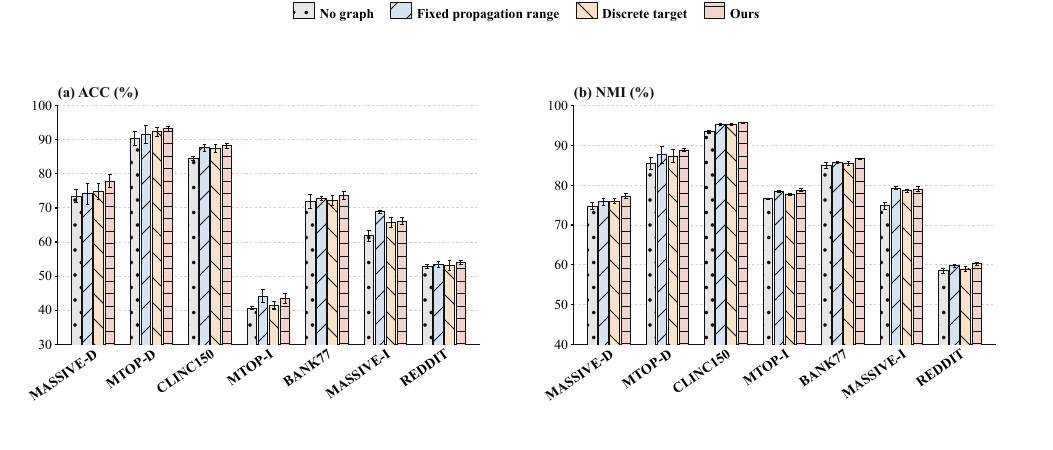}
\Description{Two bar charts side by side compare the full method and the three ablated variants under E5 across seven datasets: accuracy on the left, NMI on the right. Error bars show variation across five encoder seeds.}
\caption{Ablation on E5. Bars show the mean performance for each dataset and variant; error bars show the standard deviation across five encoder seeds.}
\label{fig:ablation:e5}
\end{figure*}

\section{Query Efficiency on the Other Embeddings}
\label{app:budget}

The Pareto frontier changes with the frozen representation. \method lies on the
frontier from $1N$ under TF-IDF; it reaches the frontier at $2N$ for both metrics
under MiniLM and E5, and at $2N$ for accuracy and $3N$ for NMI under Instructor.
At $6N$, or $21.9$k requests on average, \method obtains the highest mean ACC
and NMI in all four embedder-specific comparisons. LeBoT uses $42.5$k requests
on average; its mean ACC and NMI are lower than the $6N$ point in every panel.

The budget response is largest under TF-IDF, whose Plain baseline is also the
lowest among the four representations. From $1N$ to $6N$, TF-IDF gains $13.6$
accuracy points and $12.9$ NMI points. Across the three neural embedders, the
corresponding gains range from $5.2$ to $7.3$ points for accuracy and from $5.2$
to $6.2$ points for NMI. Most of the improvement is obtained by $2N$--$3N$,
while the $6N$ operating point achieves the highest value in every panel.

\begin{figure*}[t]
\centering
\begin{minipage}[t]{0.49\textwidth}
  \centering
  \includegraphics[width=\linewidth]{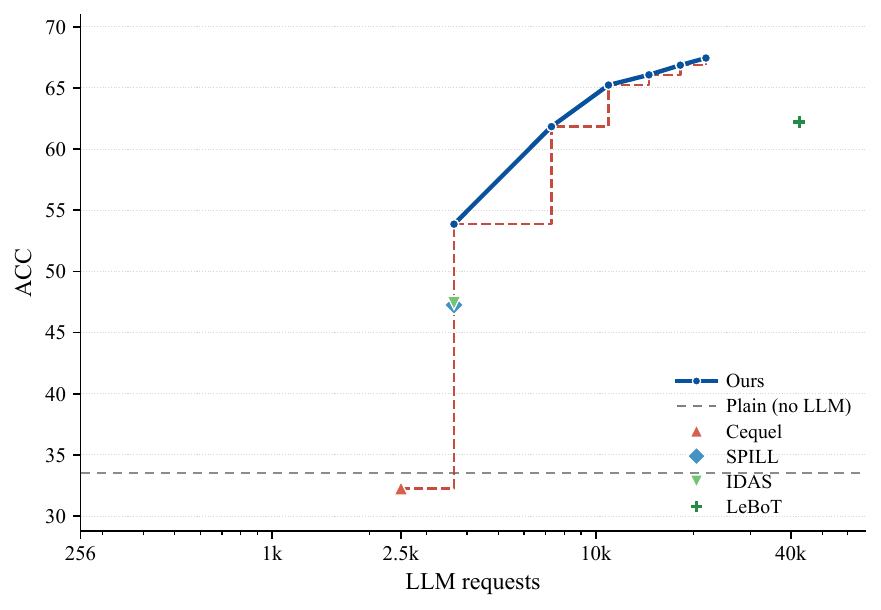}\\[-2pt]
  (a)~Accuracy
\end{minipage}\hfill
\begin{minipage}[t]{0.49\textwidth}
  \centering
  \includegraphics[width=\linewidth]{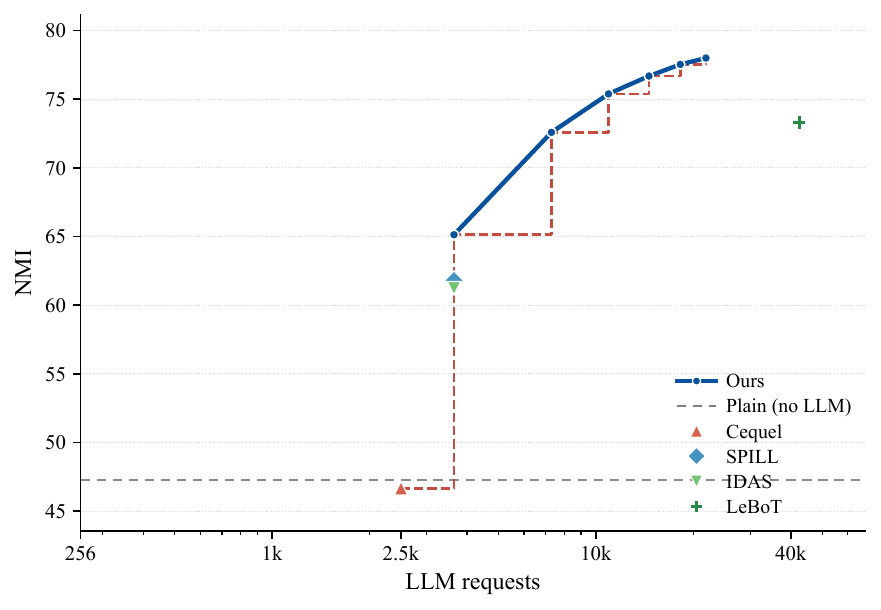}\\[-2pt]
  (b)~NMI
\end{minipage}
\Description{Two panels show mean accuracy and NMI against logarithmic LLM
request count for TF-IDF, including the \method budget curve, comparison-method
operating points, the no-LLM reference and the empirical Pareto envelope.}
\caption{Request--performance comparison with TF-IDF. \method is evaluated from
$1N$ to $6N$ requests; markers show comparison methods at their evaluated
budgets, and the red staircase shows the empirical Pareto envelope. Values are
means over seven datasets.}
\label{fig:budget:tfidf}
\end{figure*}

\begin{figure*}[t]
\centering
\begin{minipage}[t]{0.49\textwidth}
  \centering
  \includegraphics[width=\linewidth]{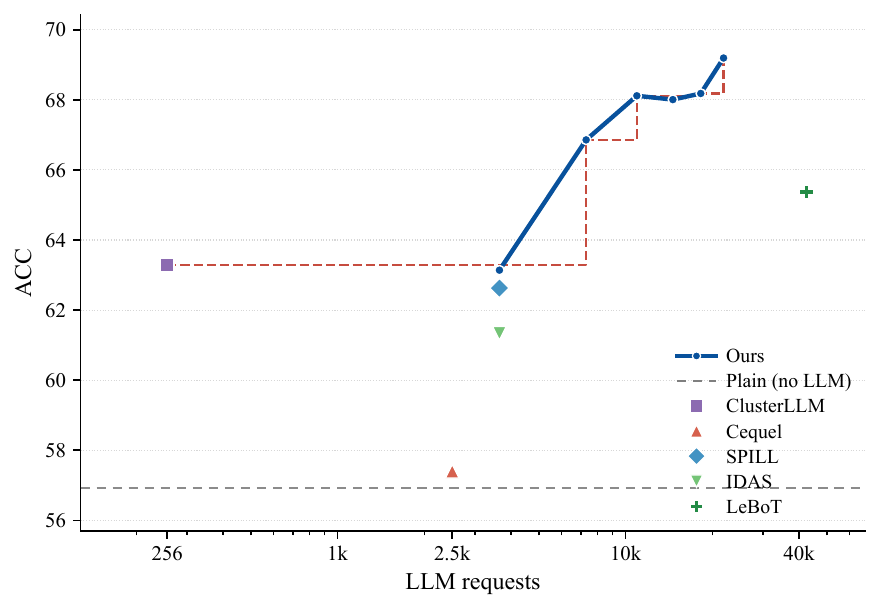}\\[-2pt]
  (a)~Accuracy
\end{minipage}\hfill
\begin{minipage}[t]{0.49\textwidth}
  \centering
  \includegraphics[width=\linewidth]{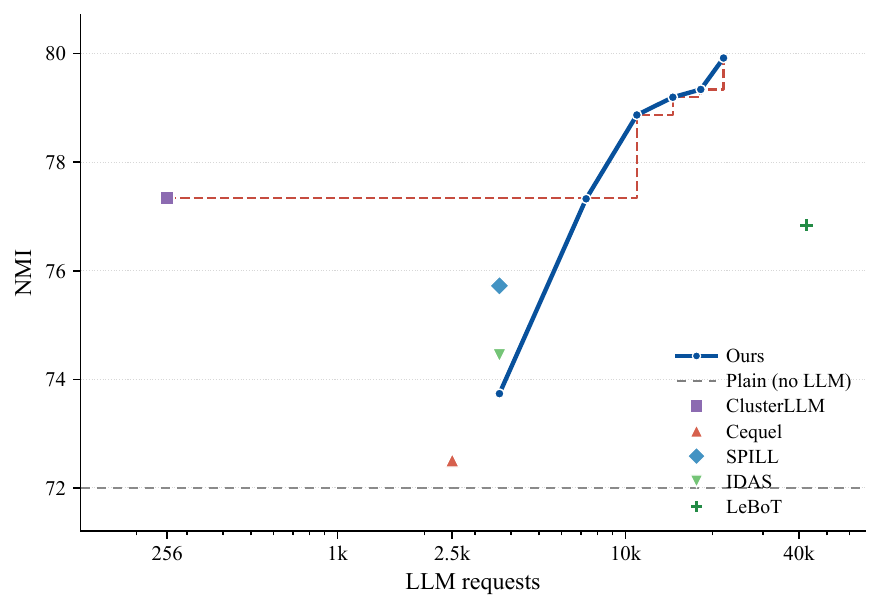}\\[-2pt]
  (b)~NMI
\end{minipage}
\Description{Two panels show mean accuracy and NMI against logarithmic LLM
request count for Instructor, including the \method budget curve,
comparison-method operating points, the no-LLM reference and the empirical
Pareto envelope.}
\caption{Request--performance comparison with Instructor. \method is evaluated
from $1N$ to $6N$ requests; markers show comparison methods at their evaluated
budgets, and the red staircase shows the empirical Pareto envelope. Values are
means over seven datasets.}
\label{fig:budget:instructor}
\end{figure*}

\begin{figure*}[t]
\centering
\begin{minipage}[t]{0.49\textwidth}
  \centering
  \includegraphics[width=\linewidth]{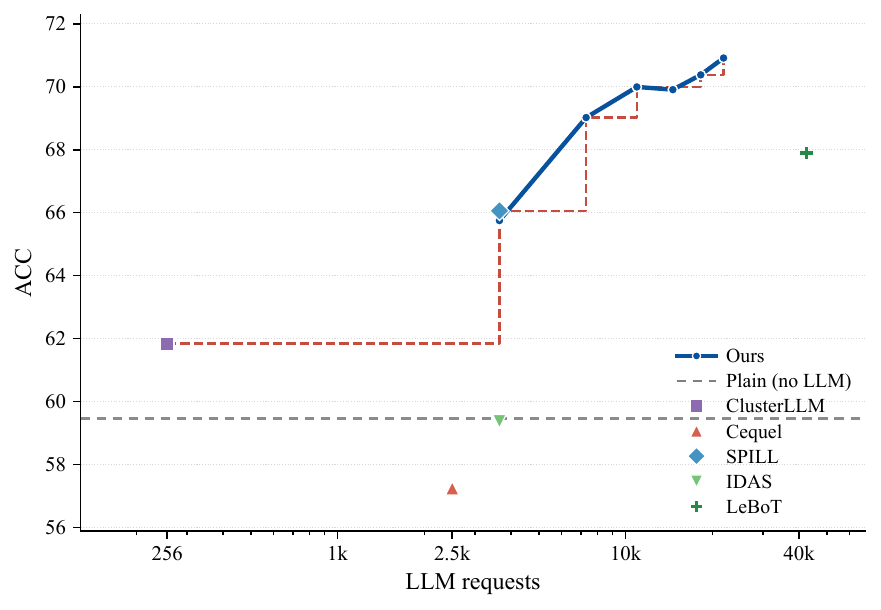}\\[-2pt]
  (a)~Accuracy
\end{minipage}\hfill
\begin{minipage}[t]{0.49\textwidth}
  \centering
  \includegraphics[width=\linewidth]{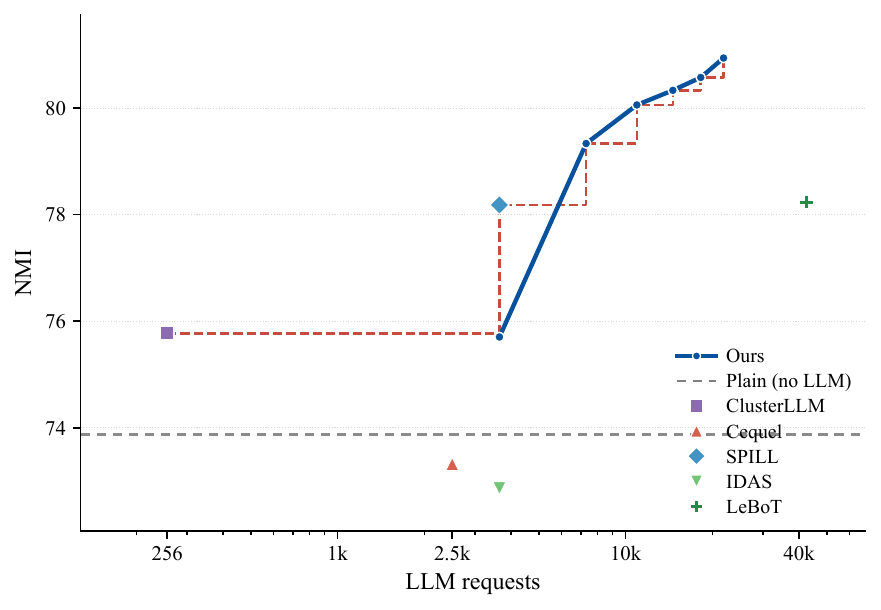}\\[-2pt]
  (b)~NMI
\end{minipage}
\Description{Two panels show mean accuracy and NMI against logarithmic LLM
request count for E5, including the \method budget curve, comparison-method
operating points, the no-LLM reference and the empirical Pareto envelope.}
\caption{Request--performance comparison with E5. \method is evaluated from
$1N$ to $6N$ requests; markers show comparison methods at their evaluated
budgets, and the red staircase shows the empirical Pareto envelope. Values are
means over seven datasets.}
\label{fig:budget:e5}
\end{figure*}

\end{document}